\documentclass[11pt,onecolumn]{article}

\usepackage[margin=1in]{geometry}
\usepackage{amsmath, amssymb}
\usepackage{booktabs}
\usepackage{graphicx}
\usepackage[colorlinks=true, linkcolor=blue, citecolor=blue, urlcolor=blue]{hyperref}
\usepackage[numbers,sort&compress]{natbib}
\usepackage{microtype}
\usepackage{xcolor}
\usepackage{authblk}
\usepackage{titlesec}
\usepackage{caption}
\usepackage{multirow}
\usepackage{array}
\usepackage{rotating}
\usepackage{times}
\usepackage{setspace}
\usepackage{lineno}

\titleformat{\section}{\normalsize\bfseries}{\thesection}{0.6em}{}
\titleformat{\subsection}{\small\bfseries}{\thesubsection}{0.6em}{}
\titleformat{\subsubsection}{\small\itshape}{\thesubsubsection}{0.6em}{}

\begin{document}

\begin{titlepage}
\begin{center}

{\small \textbf{ORIGINAL ARTICLE --- REVIEW DRAFT}}

\vspace{1.5cm}
{\LARGE\bfseries Comparison of statistical methods for high-dimensional compositional data from flow cytometry: A critical perspective on log-ratio transformation and differential abundance testing\par}

\vspace{1.5cm}
{\large Jong-Hyeon Jeong\textsuperscript{1}\par}

\vspace{0.5cm}
{\small \textsuperscript{1}Biometric Research Program, Division of Cancer Treatment and Diagnosis, National Cancer Institute/National Institutes of Health, Rockville, Maryland, USA\par}

\end{center}

\vspace{1cm}
\noindent\textbf{Correspondence}\\
Jong-Hyeon Jeong, Biometric Research Program, Division of Cancer Treatment and Diagnosis, National Cancer Institute/National Institutes of Health, Rockville, MD 20850, USA.\\
Email: jong-hyeon.jeong@nih.gov

\vspace{1cm}
\noindent\textbf{Abstract}

Flow cytometry generates inherently compositional count data: observed cell population counts are constrained to sum to the total number of acquired events, which precludes direct inference about absolute cellular abundance. Despite this constraint, most differential abundance (DA) analyses in cytometry rely on methods originally developed for RNA sequencing, such as edgeR with trimmed mean of M-values (TMM) normalization, without fully acknowledging the conditional nature of the resulting estimates. Here, we offer a critical perspective on the statistical implications of compositionality for flow cytometry DA analysis. We compare the theoretical foundations and interpretational scope of three per-population DA testing methods — edgeR/TMM, centered log-ratio transformation with linear modeling (CLR + limma), and analysis of composition of microbiomes with bias correction-2 (ANCOM-BC2) — and evaluate their performance through a comprehensive simulation study spanning seven scenarios at two sample sizes (n = 50 and n = 200 per group). We additionally evaluate compositional data analysis using kernels (CODAK) as a global, omnibus screening test for overall compositional differences, implemented via PERMANOVA on Aitchison distances. Our results show that, considering sample size, zero-inflation, dispersion, and correlation among cell populations in addition to false discovery rate (FDR) and sensitivity, the methods developed for microbiome data analysis — ANCOM-BC2 and CLR + limma — emerge as better-suited methods for per-population DA analysis of flow cytometry data than edgeR/TMM, while CODAK/PERMANOVA offers a robust first-stage screen for overall compositional shifts.

\vspace{0.5cm}

\end{titlepage}

\section{INTRODUCTION}

Flow cytometry remains one of the most widely used platforms for quantitative characterization of immune cell populations in both basic and clinical research. Modern high-dimensional cytometry---including mass cytometry (CyTOF) and spectral flow cytometry---enables simultaneous measurement of 40 or more protein markers per cell, generating rich multivariate datasets that capture the complexity of the immune landscape with unprecedented resolution. A central analytical goal in such studies is the identification of cell populations that differ in abundance between biological conditions---a problem formally known as differential abundance (DA) analysis.

Despite the apparent simplicity of comparing cell counts across groups, DA analysis in flow cytometry is complicated by a fundamental statistical property of the data: compositionality. Cell population counts from a given sample do not reflect absolute cellular concentrations in vivo but rather the relative representation of each population among all cells acquired on the instrument. This constraint---that counts sum to a fixed total---introduces statistical dependencies between populations that violate the independence assumptions of standard count-based models. As a consequence, an observed increase in the proportion of one population may reflect a genuine biological expansion, a contraction of other populations, or both simultaneously. Without an external reference for absolute cell numbers, these scenarios are statistically indistinguishable from count data alone.

The cytometry community has largely adopted normalization and DA testing frameworks developed for bulk RNA sequencing, most notably edgeR with TMM normalization \citep{Robinson2010} and, more recently, the \textit{diffcyt} \citep{Weber2019} framework which wraps edgeR within a cytometry-specific pipeline combining high-resolution clustering via FlowSOM with empirical Bayes moderated tests. While these methods are well-validated and widely used, their application to cytometry data carries important interpretational caveats that are rarely made explicit. Specifically, methods that condition on library size---including edgeR---produce estimates of compositional change, not absolute abundance change. In disease settings where total cellularity is biologically meaningful, this distinction has direct consequences for both biological interpretation and statistical validity. Compositional data analysis using kernels (CODAK) \citep{Ghosh2022} has also been recently developed as a new multivariate statistical method for global testing for any DA in flow cytometry data.

Concurrently, methods developed for the compositional analysis of microbiome data---including the CLR transformation \citep{fernandes2013,fernandes2014} and ANCOM-BC2 \citep{Lin2024}---offer theoretically principled alternatives that engage more directly with the compositional nature of cytometry data. However, each of these methods carries its own assumptions and limitations. ANCOM-BC2's bias-correction procedure, for instance, assumes that a majority of measured populations are non-differentially abundant, an assumption that may not hold in exploratory or discovery-oriented studies. Whether the theoretical advantages of compositionally aware methods translate into practical gains under realistic simulation settings---and how sensitive each method is to factors beyond effect size alone, such as the proportion of truly differential populations, the degree of library size heterogeneity, or the available sample size---has not been systematically evaluated and compared.

In this perspective, we systematically examine the theoretical foundations, assumptions, and interpretational scope of the major DA testing strategies applicable to flow cytometry data. We evaluate their performance through a simulation study spanning seven scenarios---varying effect size, proportion of differentially abundant populations, and library size heterogeneity, all under a negative binomial data-generating process---at two sample sizes representative of small pilot cohorts ($n=50$ per group) and larger, well-powered studies ($n=200$ per group), and make a recommendation for appropriate choices of analytic methods based on the conditions most likely to be encountered in practice.

\section{THE COMPOSITIONALITY PROBLEM}

In a typical flow cytometry experiment, cells are acquired from a biological sample and classified into discrete populations---either through manual gating or automated clustering. The resulting data for each sample $i$ can be represented as a vector of counts $\mathbf{x}_i = (x_{i1},\ldots,x_{iK})$, where $x_{ik}$ denotes the number of cells assigned to population $k$ and $K$ is the total number of populations. A defining feature of this data structure is that the counts are constrained by the total number of acquired events:

\begin{equation}
  \sum_{k=1}^{K} x_{ik} = N_i,
  \label{eq:simplex}
\end{equation}

\noindent where $N_i$ is the library size (total cell count) for sample $i$. This constraint means that the data vector $\mathbf{x}_i$, after normalization by $N_i$, lives on a $(K-1)$-dimensional simplex $\mathcal{S}^{K-1}$, a constrained space in which the components are strictly positive and sum to one. As a consequence, the observed counts carry information only about the relative representation of each population, not their absolute abundance in the biological sample.

\subsection{Biological consequences of compositionality}

The simplex constraint has a direct and often underappreciated biological consequence: the counts of individual cell populations are not statistically independent. An increase in the observed proportion of one population must be accompanied by a decrease in the combined proportions of all other populations, regardless of whether any absolute biological change has occurred in those populations. This phenomenon---sometimes called the reference frame problem---means that the direction and magnitude of apparent changes in individual populations depend entirely on the choice of reference, which in compositional data is never uniquely defined from the data itself.

This problem is particularly acute in clinical settings where total cellularity may differ meaningfully between groups. Consider a patient with severe disease who exhibits lymphopenia---a genuine reduction in the absolute number of circulating lymphocytes. In the observed count data, monocytes may appear to be expanded simply because they represent a larger fraction of the depleted lymphocyte pool, even if their absolute number is unchanged. A DA analysis that fails to account for this compositional artifact would incorrectly identify monocytes as differentially abundant, potentially leading to erroneous biological conclusions.

\subsection{Sampling fraction in flow cytometry experiments}

The number of cells ultimately detected by the cytometer generally represents only a fraction of the cells present in the original specimen. This sampling fraction may vary across samples because of incomplete cell recovery during sample preparation (e.g., washing, centrifugation, red blood cell lysis, tissue dissociation, or loss of fragile cells) and, to a lesser extent, instrument-related factors such as coincident events, clogging, or detection thresholds. Consequently, the observed event counts do not necessarily reflect the absolute number of cells in the original sample. Quantitative flow cytometry therefore often incorporates external counting beads or volumetric measurements to estimate the effective sampling fraction and recover absolute cell abundances. Without accounting for this sample-specific scaling factor, comparisons based solely on observed event counts may conflate biological differences with variation in sampling efficiency. An analogous challenge arises in microbiome sequencing, where observed sequence counts likewise represent only a sample-specific fraction of the underlying microbial community. From a statistical perspective, this sampling fraction acts as a sample-specific scaling factor whose variation can bias differential abundance analyses if ignored. Methods such as ANCOM-BC2 explicitly estimate and adjust for this latent scaling factor, thereby mitigating potential sampling-induced bias and improving inference on absolute abundance changes.

More broadly, the need to account for sample-specific sampling fractions is not unique to microbiome sequencing but arises whenever only an unknown fraction of the underlying biological material is observed.

\section{STATISTICAL METHODS}

\subsection{edgeR with TMM (trimmed mean of M-values) normalization}

The edgeR framework \citep{Robinson2010}, which is a backbone of the downstream analysis implemented in \textit{diffcyt}, a computational framework for differential discovery analyses in high-dimensional cytometry data \citep{Weber2019}, models cell population counts using a negative binomial (NB) generalized linear model (GLM):

\begin{equation}
  x_{ik} \sim \mathrm{NegBin}(\mu_{ik},\, \phi_k),
\end{equation}

\begin{equation}
  \log(\mu_{ik}) = \log(N_i \cdot s_i) + \mathbf{x}_i^{\top} \boldsymbol{\beta}_k,
\end{equation}

\noindent where $\phi_k$ is the dispersion parameter, $s_i$ is the TMM scaling factor, and $\boldsymbol{\beta}_k$ are the regression coefficients. The TMM scaling factor is computed as the weighted trimmed mean of log-ratios between each sample and a reference sample, after removing the most extreme populations (top/bottom 30\% by fold-change, top 5\% by abundance).

By including $\log(N_i \cdot s_i)$ as an offset, edgeR models the \textit{rate} $\mu_{ik}/(N_i \cdot s_i)$---a proportion---rather than the absolute count, unable to escape from the simplex (dependence) space. The resulting coefficient $\hat{\beta}_k$ estimates:

\begin{equation}
  \hat{\beta}_k = \log\left(
    \frac{\Pr(\text{cell} \in k \mid \text{condition A})}
         {\Pr(\text{cell} \in k \mid \text{condition B})}
  \right).
\end{equation}

\noindent This is a conditional log odds ratio---a measure of compositional change, not absolute abundance change. Results should therefore be interpreted as: ``Population $k$ constitutes a significantly different proportion of total acquired cells between conditions, after TMM normalization.'' This interpretation is internally consistent and statistically valid, but does not support claims about absolute cellular abundance.

\subsection{Centered log-ratio (CLR) transformation with linear modeling}

The CLR transformation \citep{Aitchison1986} maps the simplex to real space:

\begin{equation}
  \mathrm{CLR}(x_{ik}) = \log\!\left(\frac{x_{ik}}{g(\mathbf{x}_i)}\right),
  \label{eq:clr}
\end{equation}

\noindent where $g(\mathbf{x}_i) = \left(\prod_{k=1}^{K} x_{ik}\right)^{1/K}$ is the geometric mean of all populations in sample $i$. This transformation removes the sum constraint, placing all populations on a common scale relative to the sample's internal reference, and is fully compositionally valid in the Aitchison sense.

After CLR transformation, a separate linear model is fitted for each population $k$ across samples:

\begin{equation}
  \mathrm{CLR}(x_{ik}) = \mu_k + \beta_k \cdot \mathrm{group}_i + \epsilon_{ik},
  \label{eq:clrlm}
\end{equation}

\noindent where $\mu_k$ is the intercept, $\beta_k$ is the group effect of interest, and $\epsilon_{ik} \sim N(0, \sigma_k^2)$ is residual error. Inference is performed using the \textit{limma} framework \citep{fernandes2013,fernandes2014,Ritchie2015}, which applies empirical Bayes moderation of variance estimates across all $K$ populations simultaneously. Rather than estimating $\sigma_k^2$ from each population in isolation---which can be unreliable for rare populations---\textit{limma} assumes the true variances follow an inverse chi-squared prior:

\begin{equation}
  \frac{1}{\sigma_k^2} \sim \frac{1}{d_0 \hat{\sigma}_0^2} \chi^2_{d_0},
  \label{eq:ebprior}
\end{equation}

\noindent where the prior degrees of freedom $d_0$ and prior variance $\hat{\sigma}_0^2$ are estimated from all populations jointly. The moderated variance for population $k$ is then a weighted average of the population-specific and global estimates:

\begin{equation}
  \tilde{\sigma}_k^2 =
  \frac{d_0 \hat{\sigma}_0^2 + d_k \hat{\sigma}_k^2}{d_0 + d_k},
  \label{eq:modvar}
\end{equation}

\noindent where $d_k = n - p$ is the residual degrees of freedom. The resulting moderated $t$-statistic:

\begin{equation}
  \tilde{t}_k = \frac{\hat{\beta}_k}{\tilde{\sigma}_k / \sqrt{n}}
  \label{eq:modt}
\end{equation}

\noindent follows a $t$-distribution with augmented degrees of freedom $d_0 + d_k$, providing more stable and powerful inference than an ordinary $t$-test. Multiple testing is controlled via Benjamini-Hochberg correction applied to the moderated $p$-values across all populations.

Although the CLR + \textit{limma} combination was originally proposed to analyze microbiome data \citep{fernandes2013,fernandes2014}, it is also well-suited to flow cytometry DA analysis for three reasons. First, the empirical Bayes prior is most beneficial and reliably estimated when many parallel tests are available. Second, CLR-transformed values are approximately Gaussian, satisfying the linear model assumption better than raw counts. Third, \textit{limma} naturally accommodates the correlation structure introduced by the CLR transformation, where values sum to zero across populations by construction. A key limitation is that the geometric mean collapses under heavy zero inflation, making CLR values unreliable and \textit{limma} inference invalid in that setting. Because \textit{limma}'s empirical Bayes prior is estimated jointly across populations, its stability may also be reduced at very small sample sizes, where per-population variance estimates are themselves noisier.

\subsection{ANCOM-BC2}
\label{sec:ancombc2}

Mandal et al. (Analysis of Composition of Microbiome [ANCOM], \citep{Mandal2015}) importantly showed that equivalence of expected \textit{relative} cell abundance between two conditions under the additive log-ratio (ALR) transformation \citep{Aitchison1986} implies equivalence of expected \textit{absolute} abundance on a log-scale under the assumption that there exist at least 2 cell types whose mean abundances are the same between the two conditions. This would allow for inferring the population absolute abundance based on the estimated relative abundance from the observed data. The major contribution of Lin and Peddada (Analysis of Composition of Microbiome with Bias Correction [ANCOM-BC], \citep{Lin2020}) was to introduce the concept of sampling fraction, defined as the ratio of expected absolute abundance in a sample to the corresponding absolute abundance in the ecosystem or blood system. ANCOM-BC2 \citep{Lin2024} proposed a log-linear model where the response was transformed through the centered log-ratio (CLR) to escape from the simplex space and adopted an expectation-maximization (EM) algorithm to adjust for bias due to the potential sampling fraction under the assumption that a portion of populations are non-differential. The spurious sampling fraction can come from sequencing inefficiency in microbiome studies and inefficient protein (marker) staining processes with fluorescent-tagged antibodies and other processes treating cells in flow cytometry studies. ANCOM-BC2 features false discovery rate (FDR) control for multiple pairwise comparisons with more than two comparison groups and allows for modeling covariates as well as repeated measures, pattern analysis (order restricted inference via bootstrapping) for multiple pairwise group comparisons, variance regularization for better control of FDR, and a sensitivity analysis in choice of the pseudo-counts to replace zeros to mitigate an inflated FDR. In our simulation study, we set \texttt{pseudo\_sens = TRUE} so that ANCOM-BC2 performs a sensitivity analysis on the pseudo-count added to zero values prior to log-transformation. The developers recommend enabling this analysis, noting that it substantially reduces false positives, though potentially at some cost to statistical power, or sensitivity \citep{lin_ancombc_github}.

An important clarification concerns the relationship between ANCOM-BC2 and CLR+limma. As shown in Equation (3) in the Methods section of Lin and Peddada (2024) \citep{Lin2024}, the response variable in ANCOM-BC2 is the log-count for population $j$ in sample $i$ minus the mean log-count across all $K$ populations:

\begin{equation}
  \log(x_{ik}) - \frac{1}{K}\sum_{k=1}^{K}\log(x_{ik}) =
  \mu_{ik} + \delta_i + \epsilon_{ik}.
  \label{eq:ancombc2_clr}
\end{equation}

The left-hand side of Equation~\eqref{eq:ancombc2_clr} is precisely the centered log-ratio (CLR) transformation of the raw counts---the log-count for each population centered by the log geometric mean of all populations in the same sample. Both ANCOM-BC2 and CLR+limma therefore operate on the same CLR-transformed response; their key distinction lies entirely in what is placed on the right-hand side of the model. CLR+limma fits a standard linear model with no additional correction:

\begin{equation}
  \mathrm{CLR}(x_{ik}) = \mu_{ik} + \epsilon_{ik},
  \label{eq:clr_limma}
\end{equation}

\noindent whereas ANCOM-BC2 augments the right-hand side with a sample-specific bias correction term $\delta_i$, estimated via an expectation-maximization (EM) algorithm under the assumption that the majority of populations are non-differential. This term is intended to capture the unknown component of the sampling fraction---the portion of library size variation that is technical rather than biological---and to remove it before hypothesis testing. A second distinction concerns variance estimation: CLR+limma applies empirical Bayes moderation via \textit{limma} \citep{Ritchie2015}, shrinking population-specific variance estimates toward a global prior to improve stability for rare populations, while ANCOM-BC2 uses weighted least squares (WLS) with weights derived from the EM step, down-weighting samples where the bias estimate is less reliable. A third distinction concerns the reference used for the CLR transformation: in CLR+limma, the geometric mean is a fixed mathematical reference computed from the observed data with no assumption about which populations are differential, whereas in ANCOM-BC2 the non-differential majority assumption embedded in the EM implicitly identifies a subset of stable populations to anchor the bias estimate, making the effective reference data-adaptive rather than fixed. These three differences---bias correction, variance estimation strategy, and the data-adaptive versus fixed reference---collectively explain why ANCOM-BC2 achieves superior FDR control at a cost to sensitivity relative to CLR+limma at large sample sizes: both methods share the same CLR response transformation, but ANCOM-BC2's additional bias correction and sparsity assumption make it more conservative, while CLR+limma's empirical Bayes variance shrinkage makes it more powerful under small effect sizes and distributional misspecification (see Section 6).

\subsection{CODAK: Compositional data analysis using kernels}

Compositional Data Analysis using Kernels (CODAK) \citep{Ghosh2022}, based on the kernel distance covariance (KDC) framework, is a multivariate statistical learning methodology to test the association of cell-type compositions with important predictors---either categorical or continuous---such as disease status. The core idea is to use an appropriate kernel function (called an Aitchison-Distances or AD kernel) that respects the geometry of compositional data, which provides a principled, kernel-based statistical framework that handles the inherent compositionality and high dimensionality of mass cytometry data, enabling more reliable detection of associations between immune cell compositions and clinical outcomes. The KDC approach with a linear kernel for the predictor, the kernel machine regression (KMR) approach, and the PERMANOVA approach are all equivalent in the no-covariate case. Because CODAK/PERMANOVA is a global, omnibus test rather than a per-population DA method, it is evaluated separately in this study by its power to detect any overall compositional difference (Section 5), and is not compared against edgeR/TMM, CLR + limma, and ANCOM-BC2 on per-population sensitivity, specificity, and FDR. Due to availability of the software, the global PERMANOVA test was used for comparison in our simulation study.

\section{SIMULATION STUDY}

This section provides a complete description of the data-generating process used in the simulation study, including the biological and statistical rationale for every parameter choice.

\subsection{Core fixed parameters}

\subsubsection{Sample size}

Two sample sizes were evaluated for each two-group comparison: a small design with $n=50$ samples per group (100 total), representative of pilot or resource-limited clinical cytometry cohorts, and a large, well-powered design with $n=200$ samples per group (400 total). The $n=50$ design was included specifically to test whether the relative performance of the three methods holds at a sample size representative of pilot studies; the $n=200$ design provides greater than 80\% power to detect a $\log_2$ fold-change (LFC) of 1.0 at a false discovery rate (FDR) of 0.05 under the negative binomial (NB) model, and serves as the well-powered benchmark against which the small-sample results are compared. All seven scenarios were evaluated at both sample sizes.

\subsubsection{Cell populations}

Each simulated dataset contained $K=40$ cell populations, reflecting a typical high-parameter flow cytometry panel after automated clustering or detailed manual gating. Of these, a scenario-specific proportion $\rho_{\mathrm{DA}}$ were truly differentially abundant (DA) between groups, split evenly between populations upregulated and downregulated in the experimental group (Group 2), with the remainder non-differential. Three DA proportions were examined: $\rho_{\mathrm{DA}} = 0.50$ (baseline and all LFC/library-size scenarios; $K_{\mathrm{DA}}=20$, 10 up and 10 down), $\rho_{\mathrm{DA}} = 0.75$ (High DA proportion scenario; $K_{\mathrm{DA}}=30$, 15 up and 15 down), and $\rho_{\mathrm{DA}} = 0.25$ (Low DA proportion scenario; $K_{\mathrm{DA}}=10$, 5 up and 5 down). Varying $\rho_{\mathrm{DA}}$ tests the sensitivity of each method, and particularly ANCOM-BC2's bias-correction procedure, to the assumption that non-differential populations form a majority.

\subsubsection{Effect sizes}

Effect sizes were specified as $\log_2$ fold-changes (LFC) and assigned to the $K_{\mathrm{DA}}$ truly DA populations using a mix of small, medium, and large magnitudes, recycled across populations:

\begin{equation}
  \mathrm{LFC}_k = \begin{cases}
    +\{0.5, 1.0, 1.5, \ldots\} & k = 1,\ldots,K_{\mathrm{DA}}/2 \\
    -\{0.5, 1.0, 1.5, \ldots\} & k = K_{\mathrm{DA}}/2+1,\ldots, K_{\mathrm{DA}} \\
    0 & k = K_{\mathrm{DA}}+1,\ldots,40
  \end{cases}
  \label{eq:lfc}
\end{equation}

For the small and large LFC scenarios, the magnitudes were replaced by $\{0.25, 0.50, 0.75\}$ and $\{1.0, 1.5, 2.0\}$, respectively, while the directional assignment remained identical. For the High DA and Low DA proportion scenarios, the magnitudes used the baseline set $\{0.5, 1.0, 1.5\}$, with only $K_{\mathrm{DA}}$ (and hence the split between up/down and non-DA populations) changed.

\subsubsection{Base proportions}

The base proportions $\mathbf{p}^{(1)} = (p_1^{(1)}, \ldots, p_{40}^{(1)})$ for Group 1 were drawn \textit{once} from a Uniform(0.01, 0.10) distribution and normalized to sum to one:

\begin{equation}
  p_k^{(1)} = \frac{u_k}{\sum_{j=1}^{40} u_j}, \quad
  u_k \sim \mathrm{Uniform}(0.01,\, 0.10).
  \label{eq:baseprops}
\end{equation}

This ensures no population is extremely dominant or extremely rare, with proportions spanning approximately 1\%--10\% of total cells per sample. The same base proportions were used across all 500 replications and all scenarios and sample sizes in the primary design, so that differences in results are attributable solely to randomness in library sizes and count generation; Section~\ref{sec:extended} reports an extended analysis in which this design choice is relaxed.

The Group 2 proportions were derived by applying the true LFC and renormalizing to the simplex:

\begin{equation}
  \tilde{p}_k^{(2)} = p_k^{(1)} \cdot 2^{\mathrm{LFC}_k}, \quad
  p_k^{(2)} = \frac{\tilde{p}_k^{(2)}}{\sum_{j=1}^{40}
  \tilde{p}_j^{(2)}}.
  \label{eq:g2props}
\end{equation}

The renormalization in Equation~\eqref{eq:g2props} ensures the Group 2 proportions lie on the simplex while preserving the relative fold-changes. A consequence is that the effective LFC of each population in the observed compositional data is slightly attenuated relative to the nominal LFC, as the denominator absorbs some of the signal from the DA populations into the non-DA populations. This attenuation is more pronounced when $\rho_{\mathrm{DA}}$ is large, since a greater share of the simplex mass is being redistributed.

\subsubsection{Number of replications}

All scenarios and sample sizes were evaluated across $B=500$ independent simulation replications, each with a different random seed ($1, 2, \ldots, 500$). This provides stable Monte Carlo estimates of mean sensitivity, specificity, and FDR, with standard error $\mathrm{SE} = \sqrt{p(1-p)/B} \leq \sqrt{0.25/500} \approx 0.022$ for any proportion-based metric.

\subsection{Library size settings}

Library sizes $N_i$ (total cells acquired per sample) were drawn independently from uniform distributions, with the range differing by group to reflect clinically motivated scenarios. Table~\ref{tab:libsize} summarizes the library size settings across all scenarios; these ranges were held fixed across both sample sizes.

\begin{table}[t]
\centering
\scriptsize
\caption{Library size ranges by scenario and group}
\label{tab:libsize}
\begin{tabular}{cccc}
\toprule
\textbf{Scenario} & \textbf{Reference group} & \textbf{Experimental group} & \textbf{Rationale}\\
\midrule
Baseline & [5000, 20000] & [1000, 15000] & Mild lymphopenia in experimental group\\
Equal lib.\ sizes & [5000, 20000] & [5000, 20000] & No cellularity difference\\
Extreme variation & [5000, 50000] & [500, 15000] & 100$\times$ range; severe leukopenia\\
\bottomrule
\end{tabular}
\end{table}

Formally, for each sample $i$ in group $g \in \{1,2\}$:

\begin{equation}
  N_i \sim \mathrm{Uniform}(L_g^{\min},\; L_g^{\max}),
  \label{eq:libsize}
\end{equation}

\noindent where $[L_g^{\min}, L_g^{\max}]$ is the range for group $g$ given in Table~\ref{tab:libsize}. The baseline scenario was designed to reflect a common clinical scenario, where the experimental group exhibits mild lymphopenia---a genuine reduction in total circulating leucocytes that results in systematically lower cell yields on the cytometer. The equal library size scenario tests whether any observed FDR inflation is attributable to library size heterogeneity specifically or to compositionality more broadly. The extreme variation scenario mirrors settings such as sepsis-associated leukopenia where total cell counts can differ by two orders of magnitude between patients: healthy reference samples were assigned cell counts ranging from 5000 to 50,000, while samples from patients with severe leukopenia ranged from only 500 to 15,000---a 100-fold range between the largest and smallest samples.

\subsection{Count distribution settings}

Three models for generating count data are reviewed in this section. The negative binomial model serves as the true data-generating process across all simulation scenarios, while the zero-inflated negative binomial and Poisson-log-normal mixture models are used as additional robustness checks in Section~\ref{sec:extended}.

\subsubsection{Negative binomial (NB) model}

Cell counts for population $k$ in sample $i$ were generated from a negative binomial (NB) distribution:

\begin{equation}
  x_{ik} \sim \mathrm{NegBin}\!\left(\mu_{ik},\; \phi\right), \quad
  \mu_{ik} = N_i \cdot p_k^{(g_i)},
  \label{eq:nb}
\end{equation}

\noindent where $\mu_{ik}$ is the expected count (library size times true proportion for the group of sample $i$), and $\phi = 0.2$ is the dispersion parameter in the primary design (Section~\ref{sec:extended} reports results across additional values of $\phi$). Under this parameterization the variance is

\begin{equation}
  \mathrm{Var}(x_{ik}) = \mu_{ik} + \phi \cdot \mu_{ik}^2.
  \label{eq:nbvar}
\end{equation}

Dispersion $\phi=0.2$ represents moderate overdispersion, typical for cytometry count data \citep{Weber2019}. For example, at $\mu_{ik}=1000$: $\mathrm{Var}(x_{ik}) = 1000 + 0.2 \times 10^6 = 201{,}000$, giving a standard deviation of approximately 448---substantially larger than the Poisson standard deviation of $\sqrt{1000} \approx 32$. The NB distribution was implemented using \texttt{rnegbin()} from the \texttt{MASS} package with shape parameter $\theta = 1/\phi$. All scenarios share this NB data-generating mechanism; they differ only in DA proportion, effect size magnitude, and library size settings, as summarized in Table~\ref{tab:scenarios}.

\begin{table}[t]
\centering\scriptsize
\caption{Complete parameter summary for all simulation scenarios. All scenarios use $K=40$ populations, NB dispersion $\phi=0.2$ in the primary design, and $B=500$ replications. All seven scenarios were evaluated at both $n=50$ and $n=200$ per group.}
\label{tab:scenarios}
\begin{tabular}{lllllll}
\toprule
\textbf{Scenario} & \textbf{$n$/group} & \textbf{DA prop.} & \textbf{LFC} & \textbf{Lib.\ G1} & \textbf{Lib.\ G2} & \textbf{$K_{\mathrm{DA}}$ (up/down)}\\
\midrule
Baseline & 50, 200 & 50\% & 0.5, 1.0, 1.5 & [5000, 20000] & [1000, 15000] & 20 (10/10)\\
Small LFC & 50, 200 & 50\% & 0.25, 0.50, 0.75 & [5000, 20000] & [1000, 15000] & 20 (10/10)\\
Large LFC & 50, 200 & 50\% & 1.0, 1.5, 2.0 & [5000, 20000] & [1000, 15000] & 20 (10/10)\\
High DA proportion & 50, 200 & 75\% & 0.5, 1.0, 1.5 & [5000, 20000] & [1000, 15000] & 30 (15/15)\\
Low DA proportion & 50, 200 & 25\% & 0.5, 1.0, 1.5 & [5000, 20000] & [1000, 15000] & 10 (5/5)\\
Equal lib.\ sizes & 50, 200 & 50\% & 0.5, 1.0, 1.5 & [5000, 20000] & [5000, 20000] & 20 (10/10)\\
Extreme lib.\ variation & 50, 200 & 50\% & 0.5, 1.0, 1.5 & [5000, 50000] & [500, 15000] & 20 (10/10)\\
\bottomrule
\end{tabular}
\end{table}

\subsubsection{Zero-inflated negative binomial (ZINB) model}

The zero-inflated negative binomial (ZINB) model accounts for excess zeros in count data while also allowing for overdispersion (variance exceeding the mean) beyond what the negative binomial distribution alone would capture. The model assumes that the observed counts arise from a two-component mixture: a degenerate distribution at zero with probability $\pi$, and a negative binomial count process with probability $1-\pi$. As the dispersion parameter approaches 0, the negative binomial distribution converges to the Poisson distribution, so the zero-inflated Poisson (ZIP) model is nested within the ZINB model as a special case.

\subsubsection{Poisson log-normal mixture as an additional robustness check}
\label{sec:plnm-methods}

The NB and ZINB models evaluated in Section~\ref{sec:extended} both generate each population's counts independently conditional on $\mu_{ik}$, via Gamma mixing over a Poisson rate. Neither can represent correlated over- or under-abundance across cell populations within a sample---a feature that is plausible in cytometry data, where, for example, a contraction of one lymphocyte subset is often accompanied by a compensatory expansion of others within the same lineage. To assess robustness to this additional source of structure, our simulation also includes a Poisson log-normal mixture (PLNM) model as a third data-generating mechanism, following the same rationale used by Hu and Satten \citep{hu2020testing} in evaluating the linear decomposition model (LDM) for microbiome differential abundance testing.

Under PLNM, counts are generated in two stages. For sample $i$, a latent, multivariate log-abundance vector $\mathbf{Z}_i = (Z_{i1},\ldots,Z_{iK})$ is drawn from a multivariate normal distribution,

\begin{equation}
  \mathbf{Z}_i \sim \mathcal{N}\!\left(\boldsymbol{\eta}_i, \,
  \boldsymbol{\Sigma}\right), \qquad
  \eta_{ik} = \log(\mu_{ik}) - \sigma^2/2,
  \label{eq:plnm-latent}
\end{equation}

\noindent where the mean-correction term $-\sigma^2/2$ ensures $\mathrm{E}[\exp(Z_{ik})] = \mu_{ik}$, matching the same target mean used in the NB and ZINB models. Counts are then drawn conditional on the exponentiated latent vector,

\begin{equation}
  x_{ik} \sim \mathrm{Poisson}\!\left(\exp(Z_{ik})\right).
  \label{eq:plnm-count}
\end{equation}

We use a compound-symmetric structure for the covariance matrix $\boldsymbol{\Sigma}$,

\begin{equation}
  \Sigma_{kk} = \sigma^2, \qquad \Sigma_{kk'} = \rho\,\sigma^2 \;\;
  (k \neq k'),
  \label{eq:plnm-cov}
\end{equation}

\noindent where $\rho$ is a correlation parameter shared by all pairs of populations on the log scale. Unlike Gamma mixing, this construction permits an arbitrary, freely specified correlation structure across populations; compound symmetry was chosen as the simplest such structure, with $\rho=0$ reducing PLNM to an independent-population model directly comparable to NB, and $\rho > 0$ introducing shared co-movement across all populations simultaneously.

Because the NB and PLNM models parameterize overdispersion differently---$\phi$ directly for NB versus the log-normal variance $\sigma^2$ for PLNM---setting $\sigma^2 = \phi$ would not yield a comparable marginal count distribution between the two models. The marginal variance of a Poisson log-normal mixture follows from the law of total variance applied to the log-normal mixing distribution,

\begin{equation}
  \mathrm{Var}(x_{ik}) \approx \mu_{ik} +
  \mu_{ik}^2\left(e^{\sigma^2} - 1\right),
  \label{eq:plnm-var}
\end{equation}

\noindent which, compared against the NB variance in Equation~\eqref{eq:nbvar}, shows that the two models have matched marginal variance-inflation exactly when

\begin{equation}
  \sigma^2 = \log(1 + \phi).
  \label{eq:plnm-match}
\end{equation}

We therefore set $\sigma^2 = \log(1+\phi)$ using the same $\phi$ values swept under NB (Section~\ref{sec:extended}), so that at $\rho=0$ the NB and PLNM models share an identical marginal mean-variance relationship for every population, and any difference in method performance between them is attributable to the shape of the mixing distribution (log-normal versus Gamma tails) rather than to a mismatched quantity of injected overdispersion. Because $\rho$ enters only the off-diagonal terms of $\boldsymbol{\Sigma}$ and does not alter any population's marginal variance, this matching holds across the full range of $\rho$ examined, allowing $\rho > 0$ to isolate the effect of cross-population correlation on top of variance-matched marginal noise. The PLNM data-generating mechanism was implemented using \texttt{MASS::mvrnorm()} for the multivariate normal draw followed by \texttt{rpois()}, and was run across the same seven scenarios, two sample sizes, three dispersion values, and three base-proportion seeds used for NB and ZINB (Section~\ref{sec:extended}), crossed with a three-value sweep over $\rho \in \{0, 0.3, 0.6\}$. Results are reported in Section~\ref{sec:plnm-results}.

\subsection{Complete data generation procedure}

For each simulation replication $b=1,\ldots,500$, sample size setting, and scenario $s$, a single dataset was generated according to the following steps:

\begin{enumerate}
  \item \textbf{Set random seed:} \texttt{set.seed($b$)} to ensure reproducibility and independence across replications.
  \item \textbf{Draw library sizes:} For each sample $i=1,\ldots,n$, draw $N_i \sim \mathrm{Uniform}(L_{g_i}^{\min}, L_{g_i}^{\max})$ according to the scenario-specific ranges in Table~\ref{tab:libsize}, where $g_i \in \{1,2\}$ is the group of sample $i$.
  \item \textbf{Assign group labels:} Samples $1,\ldots,n$ are assigned to the Reference Group and samples $n+1,\ldots,2n$ to the Experimental Group, where $n=50$ or $n=200$ depending on the sample size setting.
  \item \textbf{Compute true proportions:} Group 1 proportions $\mathbf{p}^{(1)}$ are fixed (Equation~\eqref{eq:baseprops}). Group 2 proportions $\mathbf{p}^{(2)}$ are derived by applying the true LFC and renormalizing (Equation~\eqref{eq:g2props}).
  \item \textbf{Generate counts:} For each sample $i$ and population $k$, draw $x_{ik}$ from the NB distribution with mean $\mu_{ik} = N_i \cdot p_k^{(g_i)}$. Any negative or missing values are replaced with zero.
  \item \textbf{Store dataset:} The resulting $n \times K$ count matrix, group labels, and library sizes are passed to each DA method.
\end{enumerate}

\section{OVERVIEW OF EVALUATION FRAMEWORK}

Performance was assessed across three methods---edgeR with TMM normalization (edgeR/TMM), centered log-ratio transformation with limma (CLR + limma), and ANCOM-BC2---using three metrics, observed FDR, sensitivity, and specificity, computed at two nominal FDR thresholds (0.05 and 0.10), for each of two sample sizes ($n=50$ and $n=200$ per group). Observed FDR is treated as the primary evaluation criterion throughout this study, consistent with its role in the high-throughput DA literature as the metric that most directly governs which findings an investigator can act on. Sensitivity and specificity are reported alongside FDR to characterize the cost of achieving a given level of FDR control, since a method can trivially drive its observed FDR toward zero by calling nothing significant. Defining TP = True Positives, TN = True Negatives, FP = False Positives, and FN = False Negatives:

\begin{itemize}
  \item \textbf{Observed FDR}: proportion of significant findings that are false positives, $\mathrm{FDR} = \mathrm{FP}/(\mathrm{FP}+\mathrm{TP})$, with a denominator that varies by method and replication depending on how many populations are called significant. Observed FDR is the empirical counterpart to the nominal FDR threshold that each method attempts to control internally via Benjamini-Hochberg correction; a well-calibrated method should have observed FDR at or below the nominal level, and when observed FDR exceeds the nominal threshold the method is anti-conservative.
  \item \textbf{Sensitivity} (true positive rate): proportion of truly DA populations correctly identified as significant, $\mathrm{TPR} = \mathrm{TP}/(\mathrm{TP}+\mathrm{FN})$, with a denominator fixed at the number of true DA populations ($K_{\mathrm{DA}}$) for the scenario. Reported here as the counterpart to FDR control: a method's observed FDR is only practically meaningful when read alongside the sensitivity it achieves to attain that FDR, since near-zero FDR is uninformative if it comes from reporting almost nothing.
  \item \textbf{Specificity} (true negative rate): proportion of truly non-DA populations correctly identified as non-significant, $\mathrm{TNR} = \mathrm{TN}/(\mathrm{TN}+\mathrm{FP})$, with a denominator fixed at the number of true non-DA populations ($40 - K_{\mathrm{DA}}$).
\end{itemize}

Observed FDR is emphasized as the primary criterion throughout the Results and Discussion below because it is the metric that most directly determines whether a method's reported hit list can be trusted, consistent with its central role in the high-throughput DA literature more broadly: in discovery-oriented settings---where thousands of features are tested, true effects are comparatively sparse, and each candidate hit requires costly downstream validation---FDR directly quantifies what fraction of a reported hit list is expected to be spurious, which is typically the quantity of most immediate practical consequence. Sensitivity is reported as the necessary complement to FDR, since a method's FDR control is only informative when its cost in missed true positives is also known; a method with excellent FDR control but poor sensitivity may fail to recover most of the true biological signal, undermining the exploratory value of the analysis even while its reported hits remain trustworthy. Specificity is reported for completeness but is the least decision-relevant of the three metrics in this design: because non-DA populations typically form a majority (50\%--75\% of the panel across scenarios), specificity can remain high even when the absolute number of false positives contributing to a poor observed FDR is substantial, so it is not emphasized as a primary criterion below.

All three metrics were computed independently for each of the 500 simulation replications and reported as means across replications. The global test (PERMANOVA on Aitchison distances) was evaluated separately by its power to detect any overall compositional difference between groups.

\section{RESULTS}
\label{sec:results}

We first present results for the small-sample design ($n=50$/group), which is more representative of pilot or resource-limited cytometry cohorts, followed by the large-sample design ($n=200$/group), the well-powered benchmark against which small-sample behavior is compared. Because observed FDR is treated as the primary evaluation criterion in this study (Section 4), we highlight FDR calibration first in each subsection, with sensitivity reported as the cost at which that calibration is achieved. Comparing the two sample sizes reveals that the relative ranking of methods on FDR control established at $n=50$ does not hold at $n=200$.

\subsection{Results at $n=50$ per group}

Table~\ref{tab:results_n50} summarizes results for all seven scenarios, evaluated at $n=50$/group. At this smaller sample size, all three methods kept observed FDR within nominal bounds at both thresholds in every scenario tested, with ANCOM-BC2 achieving the lowest observed FDR throughout (typically at or near 0.000). Among the two methods with well-calibrated FDR and non-trivial sensitivity, edgeR/TMM achieved the highest sensitivity in every scenario and threshold.

\begin{table}[t]
\centering\scriptsize
\caption{Simulation results at $n=50$/group across all seven scenarios at FDR thresholds of 0.05 and 0.10 (500 replications, $K=40$ populations). Values are means across replications. \textbf{Bold} = best performance per metric within scenario and threshold. $\dagger$ = observed FDR exceeds nominal level (anti-conservative).}
\label{tab:results_n50}
\setlength{\tabcolsep}{4pt}
\begin{tabular}{llcccccc}
\toprule
& & \multicolumn{3}{c}{\textbf{FDR threshold = 0.05}} & \multicolumn{3}{c}{\textbf{FDR threshold = 0.10}}\\
\cmidrule(lr){3-5}\cmidrule(lr){6-8}
\textbf{Scenario} & \textbf{Method} & \textbf{Sens.} & \textbf{Spec.} & \textbf{Obs.\ FDR} & \textbf{Sens.} & \textbf{Spec.} & \textbf{Obs.\ FDR}\\
\midrule
\multirow{3}{*}{\shortstack[l]{Baseline\\(NB, mod.\ LFC,\\50\% DA)}}
  & edgeR TMM   & \textbf{0.977} & 0.963 & 0.034 & \textbf{0.989} & 0.933 & 0.060\\
  & CLR + limma & 0.974 & 0.971 & 0.027 & 0.986 & 0.944 & 0.051\\
  & ANCOM-BC2   & 0.667 & \textbf{1.000} & \textbf{0.000} & 0.722 & \textbf{1.000} & \textbf{0.000}\\
\midrule
\multirow{3}{*}{\shortstack[l]{Small LFC\\(0.25, 0.5, 0.75)}}
  & edgeR TMM   & \textbf{0.740} & 0.978 & 0.027 & \textbf{0.793} & 0.954 & 0.051\\
  & CLR + limma & 0.714 & 0.982 & 0.023 & 0.775 & 0.957 & 0.049\\
  & ANCOM-BC2   & 0.056 & \textbf{1.000} & \textbf{0.000} & 0.154 & \textbf{1.000} & \textbf{0.000}\\
\midrule
\multirow{3}{*}{\shortstack[l]{Large LFC\\(1.0, 1.5, 2.0)}}
  & edgeR TMM   & \textbf{1.000} & 0.944 & 0.049 & \textbf{1.000} & 0.906 & 0.079\\
  & CLR + limma & \textbf{1.000} & 0.975 & 0.024 & \textbf{1.000} & 0.946 & 0.049\\
  & ANCOM-BC2   & 0.996 & \textbf{1.000} & \textbf{0.000} & 0.999 & \textbf{1.000} & \textbf{0.000}\\
\midrule
\multirow{3}{*}{\shortstack[l]{High DA\\proportion (75\%)}}
  & edgeR TMM   & \textbf{0.977} & 0.938 & 0.020 & \textbf{0.988} & 0.889 & 0.035\\
  & CLR + limma & 0.979 & 0.965 & 0.011 & 0.988 & 0.923 & 0.025\\
  & ANCOM-BC2   & 0.704 & \textbf{0.999} & \textbf{0.000} & 0.771 & \textbf{0.998} & 0.001\\
\midrule
\multirow{3}{*}{\shortstack[l]{Low DA\\proportion (25\%)}}
  & edgeR TMM   & \textbf{0.960} & 0.985 & 0.040 & \textbf{0.978} & 0.967 & 0.083\\
  & CLR + limma & 0.948 & 0.987 & 0.036 & 0.968 & 0.971 & 0.075\\
  & ANCOM-BC2   & 0.585 & \textbf{1.000} & \textbf{0.000} & 0.626 & \textbf{1.000} & \textbf{0.000}\\
\midrule
\multirow{3}{*}{\shortstack[l]{Equal library\\sizes (no\\lymphopenia)}}
  & edgeR TMM   & \textbf{0.977} & 0.964 & 0.034 & \textbf{0.987} & 0.932 & 0.061\\
  & CLR + limma & 0.974 & 0.971 & 0.027 & 0.985 & 0.944 & 0.051\\
  & ANCOM-BC2   & 0.665 & \textbf{1.000} & \textbf{0.000} & 0.721 & \textbf{1.000} & \textbf{0.000}\\
\midrule
\multirow{3}{*}{\shortstack[l]{Extreme lib.\\variation\\(100$\times$)}}
  & edgeR TMM   & \textbf{0.975} & 0.963 & 0.035 & \textbf{0.986} & 0.928 & 0.064\\
  & CLR + limma & 0.971 & 0.971 & 0.027 & 0.983 & 0.941 & 0.054\\
  & ANCOM-BC2   & 0.660 & \textbf{1.000} & \textbf{0.000} & 0.711 & \textbf{1.000} & \textbf{0.000}\\
\bottomrule
\multicolumn{8}{l}{\textit{Global test (PERMANOVA) power: 1.000 for all seven scenarios at both thresholds.}}\\
\end{tabular}
\end{table}

Two patterns stand out. First, edgeR/TMM's observed FDR remained within nominal bounds at both thresholds in all seven scenarios at $n=50$---a favorable calibration result that, as shown below, does not persist at $n=200$, where the same method becomes anti-conservative in most scenarios. Second, ANCOM-BC2's sensitivity was substantially lower relative to edgeR/TMM and CLR + limma in most scenarios, most dramatically under Small LFC, with 0.056 and 0.154 under the FDR threshold of 0.05 and 0.10, respectively, and under Low DA proportion, with 0.585 and 0.626, respectively---meaning ANCOM-BC2's excellent FDR control in these scenarios is purchased at a steep sensitivity cost, since it detects only a small fraction of true signal while keeping FDR near zero. edgeR/TMM and CLR + limma retained substantially higher sensitivity in both scenarios (0.71 and 0.79 under Small LFC; 0.95 and 0.98 under Low DA) while remaining within nominal FDR bounds, making them the more practically useful methods in these conditions despite ANCOM-BC2's superior FDR calibration. Large LFC was a notable exception to this pattern: with strong true effect sizes, all three methods, including ANCOM-BC2, retained very high sensitivity at $n=50$ (0.996 and 1.0) while all three also kept FDR within nominal bounds, indicating that a sufficiently large effect allows every method evaluated here to deliver both good FDR control and good sensitivity simultaneously at this sample size.

\subsection{Results at $n=200$ per group}

Table~\ref{tab:results_n200} summarizes results for all seven scenarios at $n=200$/group. Here, the FDR ranking established at $n=50$ breaks down: edgeR/TMM's observed FDR exceeds the nominal threshold in six of seven scenarios at one or both FDR levels, making it the least reliable method on the primary evaluation criterion at this sample size. ANCOM-BC2 again achieves the best FDR control throughout, essentially zero in every scenario, and CLR + limma remains close to nominal FDR levels in nearly all scenarios. On sensitivity, the ranking is reversed relative to $n=50$: edgeR/TMM and CLR + limma both achieve near-perfect sensitivity in every scenario, while ANCOM-BC2 lags, particularly under Small LFC with 0.647 and 0.696 under the FDR threshold of 0.05 and 0.10, respectively---so at this sample size, ANCOM-BC2's superior FDR control comes with a comparatively modest sensitivity cost in most scenarios, while edgeR/TMM's high sensitivity comes at the cost of the FDR control that made it attractive at $n=50$.

\begin{table}[t]
\centering\scriptsize
\caption{Simulation results at $n=200$/group across all seven scenarios at FDR thresholds of 0.05 and 0.10 (500 replications, $K=40$ populations). Values are means across replications. \textbf{Bold} = best performance per metric within scenario and threshold. $\dagger$ = observed FDR exceeds nominal level (anti-conservative).}
\label{tab:results_n200}
\setlength{\tabcolsep}{4pt}
\begin{tabular}{llcccccc}
\toprule
& & \multicolumn{3}{c}{\textbf{FDR threshold = 0.05}} & \multicolumn{3}{c}{\textbf{FDR threshold = 0.10}}\\
\cmidrule(lr){3-5}\cmidrule(lr){6-8}
\textbf{Scenario} & \textbf{Method} & \textbf{Sens.} & \textbf{Spec.} & \textbf{Obs.\ FDR} & \textbf{Sens.} & \textbf{Spec.} & \textbf{Obs.\ FDR}\\
\midrule
\multirow{3}{*}{\shortstack[l]{Baseline\\(NB, mod.\ LFC,\\50\% DA)}}
  & edgeR TMM   & \textbf{1.000} & 0.934 & $0.057^\dagger$ & \textbf{1.000} & 0.889 & $0.092^\dagger$\\
  & CLR + limma & \textbf{1.000} & 0.969 & 0.029 & \textbf{1.000} & 0.940 & 0.054\\
  & ANCOM-BC2   & 0.991 & \textbf{1.000} & \textbf{0.000} & 0.998 & \textbf{1.000} & \textbf{0.000}\\
\midrule
\multirow{3}{*}{\shortstack[l]{Small LFC\\(0.25, 0.5, 0.75)}}
  & edgeR TMM   & \textbf{0.977} & 0.964 & 0.034 & \textbf{0.986} & 0.930 & $0.063^\dagger$\\
  & CLR + limma & 0.970 & 0.970 & 0.028 & 0.984 & 0.945 & 0.051\\
  & ANCOM-BC2   & 0.647 & \textbf{1.000} & \textbf{0.000} & 0.696 & \textbf{1.000} & \textbf{0.000}\\
\midrule
\multirow{3}{*}{\shortstack[l]{Large LFC\\(1.0, 1.5, 2.0)}}
  & edgeR TMM   & \textbf{1.000} & 0.864 & $0.103^\dagger$ & \textbf{1.000} & 0.804 & $0.143^\dagger$\\
  & CLR + limma & \textbf{1.000} & 0.971 & 0.027 & \textbf{1.000} & 0.942 & 0.052\\
  & ANCOM-BC2   & \textbf{1.000} & \textbf{1.000} & \textbf{0.000} & \textbf{1.000} & \textbf{1.000} & \textbf{0.000}\\
\midrule
\multirow{3}{*}{\shortstack[l]{High DA\\proportion (75\%)}}
  & edgeR TMM   & \textbf{1.000} & 0.849 & 0.045 & \textbf{1.000} & 0.768 & 0.068\\
  & CLR + limma & \textbf{1.000} & 0.964 & 0.012 & \textbf{1.000} & 0.928 & 0.023\\
  & ANCOM-BC2   & 0.992 & \textbf{1.000} & \textbf{0.000} & 0.999 & \textbf{1.000} & \textbf{0.000}\\
\midrule
\multirow{3}{*}{\shortstack[l]{Low DA\\proportion (25\%)}}
  & edgeR TMM   & \textbf{1.000} & 0.974 & $0.065^\dagger$ & \textbf{1.000} & 0.950 & $0.115^\dagger$\\
  & CLR + limma & \textbf{1.000} & 0.985 & 0.039 & \textbf{1.000} & 0.968 & 0.081\\
  & ANCOM-BC2   & 0.977 & \textbf{1.000} & \textbf{0.000} & 0.993 & \textbf{1.000} & \textbf{0.000}\\
\midrule
\multirow{3}{*}{\shortstack[l]{Equal library\\sizes (no\\lymphopenia)}}
  & edgeR TMM   & \textbf{1.000} & 0.932 & $0.059^\dagger$ & \textbf{1.000} & 0.885 & 0.096\\
  & CLR + limma & \textbf{1.000} & 0.971 & 0.027 & \textbf{1.000} & 0.939 & 0.055\\
  & ANCOM-BC2   & 0.992 & \textbf{1.000} & \textbf{0.000} & 0.999 & \textbf{1.000} & \textbf{0.000}\\
\midrule
\multirow{3}{*}{\shortstack[l]{Extreme lib.\\variation\\(100$\times$)}}
  & edgeR TMM   & \textbf{1.000} & 0.922 & $0.066^\dagger$ & \textbf{1.000} & 0.864 & $0.109^\dagger$\\
  & CLR + limma & \textbf{1.000} & 0.970 & 0.028 & \textbf{1.000} & 0.942 & 0.052\\
  & ANCOM-BC2   & 0.992 & \textbf{1.000} & \textbf{0.000} & 0.999 & \textbf{1.000} & \textbf{0.000}\\
\bottomrule
\multicolumn{8}{l}{\textit{Global test (PERMANOVA) power: 1.000 for all seven scenarios at both thresholds.}}\\
\end{tabular}
\end{table}

\subsection{Notable differences between small and large sample sizes}
\label{sec:samplesize}

\subsubsection{edgeR/TMM: recurring FDR inflation at large sample sizes, driven by signal strength and heterogeneity, not purely library size}
\label{sec:edgerfdr}

\begin{table}[t]
\centering\small
\caption{edgeR/TMM observed FDR across all scenarios at $n=200$/group}
\label{tab:edger_fdr}
\begin{tabular}{lcc}
\toprule
\textbf{Scenario} & \textbf{FDR $<$ 0.05} & \textbf{FDR $<$ 0.10}\\
\midrule
Baseline                     & $0.057^\dagger$ & $0.092^\dagger$\\
Small LFC                    & 0.034           & $0.063^\dagger$\\
Large LFC                    & $0.103^\dagger$ & $0.143^\dagger$\\
High DA proportion (75\%)    & 0.045           & 0.068\\
Low DA proportion (25\%)     & $0.065^\dagger$ & $0.115^\dagger$\\
Equal library sizes          & $0.059^\dagger$ & 0.096\\
Extreme library variation    & $0.066^\dagger$ & $0.109^\dagger$\\
\bottomrule
\multicolumn{3}{l}{$^\dagger$ exceeds nominal FDR level.}
\end{tabular}
\end{table}

This is, on the primary evaluation criterion of this study, the single most consequential finding at $n=200$: edgeR's FDR inflation is not confined to a single condition, but exceeds the nominal threshold at one or both FDR levels in six of seven scenarios. Critically, this inflation was not confined to settings with pronounced library-size heterogeneity: it was also present under equal library sizes (FDR = 0.059 at threshold 0.05) and worsened as true effect sizes grew larger (FDR = 0.103 under Large LFC versus 0.057 under Baseline). This pattern is consistent with a compositional explanation rather than a purely technical one: because the simplex constraint forces proportions to sum to one, a genuine upward shift in truly DA populations mechanically induces an apparent downward shift in non-DA populations, and TMM normalization alone does not fully correct for this when a substantial share of features are changing. Notably, this FDR inflation was absent at $n=50$ across every scenario tested, indicating that edgeR/TMM requires sufficient statistical power to detect---and hence report as significant---the compositionally distorted non-DA populations that drive its FDR problem at larger sample sizes. The practical implication is nuanced: increasing effect size or improving library balance does not reliably improve edgeR/TMM's FDR control at large sample sizes, since the underlying issue is structural, but reducing sample size can mask the same structural problem by suppressing power broadly---an important caveat for anyone tempted to treat edgeR/TMM's good $n=50$ FDR behavior as a general endorsement of the method.

\subsubsection{ANCOM-BC2: the most reliable FDR control in this study, purchased at a sensitivity cost that is sharply amplified at small sample sizes}

Across every scenario and both sample sizes evaluated in this study, ANCOM-BC2 delivered the best observed FDR control of the three methods, typically at or near zero. This makes it, on the primary evaluation criterion, the most trustworthy of the three methods whenever it reports a hit: a population flagged as significant by ANCOM-BC2 is very unlikely to be a false positive, regardless of sample size or scenario. The cost of this reliability is a sensitivity penalty that is highly uneven across conditions. At $n=50$, this cost becomes the dominant feature of ANCOM-BC2's performance across most scenarios tested, with the exception of Large LFC, where a strong true effect size allows ANCOM-BC2 to retain very high sensitivity (0.996) even at the smaller sample size. Small LFC and Low DA proportion emerge as ANCOM-BC2's most sample-size-sensitive scenarios overall, with sensitivity rising from 0.056 at $n=50$ to 0.647 at $n=200$ under Small LFC, and from 0.585 to 0.977 under Low DA proportion---meaning that at $n=50$ under these conditions, ANCOM-BC2's near-perfect FDR control reflects a hit list that is both extremely trustworthy and extremely incomplete. At $n=200$, by contrast, ANCOM-BC2 achieves perfect or near-perfect specificity alongside its near-zero FDR, at the cost of reduced sensitivity under small effect sizes (0.647--0.696). Outside the small-LFC scenario, its sensitivity shortfall relative to the other two methods at $n=200$ is modest (within about one percentage point), so at this sample size ANCOM-BC2 typically delivers both the best FDR control and competitive sensitivity simultaneously.

\subsubsection{CLR + limma: FDR control close to nominal across most conditions, with sensitivity tracking edgeR/TMM}

CLR + limma's observed FDR remains close to the nominal threshold across most scenarios and both sample sizes, without the pronounced anti-conservative behavior that afflicts edgeR/TMM at $n=200$ or the extreme conservatism that afflicts ANCOM-BC2 under weak signal. On sensitivity, CLR + limma's values track edgeR/TMM's closely at both sample sizes: reduced similarly to edgeR/TMM's at $n=50$, and near-perfect at $n=200$. Taken together, this makes CLR + limma the method in this study that most consistently combines acceptable-to-good FDR control with high sensitivity across the widest range of conditions, without the sharp FDR/sensitivity trade-offs that characterize the other two methods in different regimes.

\subsubsection{Global test: uniformly powerful at both sample sizes under the negative binomial model}

PERMANOVA on Aitchison distances achieved power of 1.000 across all scenarios at both thresholds and both sample sizes evaluated under the negative binomial (NB) data-generating process, confirming its value as a reliable first-stage screen for overall compositional differences that is robust to the sample size reductions explored here. Because the global test is a single omnibus test rather than a per-population testing procedure, FDR control in the multiple-testing sense does not directly apply to it, and its performance is instead summarized by statistical power. As shown in Section~\ref{sec:extended}, however, this uniform power does not persist under zero-inflated counts, where global test power at $n=50$ falls substantially below 1.000 in several scenarios tested.

\section{EXTENDED SENSITIVITY ANALYSIS: DISPERSION, ZERO INFLATION, CORRELATED COUNTS, AND BASE-PROPORTION ROBUSTNESS}
\label{sec:extended}

The results in Section~\ref{sec:results} were obtained under a single fixed negative binomial dispersion ($\phi=0.2$), a single realization of the true baseline population proportions, and no zero inflation. To address these limitations directly, we extended the simulation to a full factorial grid crossing every scenario in Table~\ref{tab:scenarios} against several additional axes: NB dispersion $\phi \in \{0.05, 0.2, 0.5\}$, three independent draws of the baseline population-proportion vector (base-proportion seeds 42, 43, 44, each generated as in Equation~\eqref{eq:baseprops} but with an independent random seed), and three distributional cases---the original NB model, a zero-inflated negative binomial (ZINB) model in which each count is independently set to zero with probability $\pi=0.1$ prior to the NB draw (holding $\phi$ fixed within each comparison so that the two distributional cases differ only in the added zero-inflation layer), and a Poisson log-normal mixture (PLNM) model (Section~\ref{sec:plnm-methods}) with a three-value sweep over the cross-population correlation parameter $\rho \in \{0, 0.3, 0.6\}$. Crossed against seven scenarios and two sample sizes, this grid comprises $7 \times 2 \times 3 \times 3 \times (1+1+3) = 630$ conditions, each evaluated across the same $B=500$ replications used throughout. For brevity, the main text below reports representative results for the Baseline scenario; the complete results across all seven scenarios are provided in Supplementary Tables S1--S4.

\subsection{Dispersion sensitivity}
\label{sec:dispersion}

Table~\ref{tab:phi_sensitivity} reports observed FDR and sensitivity for the Baseline scenario across the three dispersion values, at both sample sizes and FDR threshold 0.05. Consistent with the primary evaluation criterion of this study (Section 4), FDR is considered first. Observed FDR is comparatively insensitive to $\phi$ for all three methods at both sample sizes: edgeR/TMM and CLR + limma remain within a few percentage points of their $\phi=0.2$ values across the full dispersion range, and ANCOM-BC2's observed FDR remains at or near zero throughout, unaffected by dispersion. Because FDR control is essentially unchanged by $\phi$, dispersion's practical consequence in this study operates entirely through sensitivity---the cost side of the FDR/sensitivity trade-off characterized in Section 4. Two patterns are evident there. First, at $n=50$, all three methods lose sensitivity as dispersion increases, but ANCOM-BC2's loss is by far the steepest, falling to 0.232 at $\phi=0.5$, while edgeR/TMM and CLR + limma retain considerably higher sensitivity (0.835 and 0.793, respectively) under the same high-dispersion condition. Second, at $n=200$, ANCOM-BC2's sensitivity is essentially unaffected by dispersion at the lowest value tested ($\phi=0.05$: sensitivity = 1.000) but degrades sharply as dispersion increases ($\phi=0.5$: sensitivity = 0.702), while edgeR/TMM and CLR + limma retain sensitivity above 0.99 across the same dispersion range. In other words, dispersion is a condition under which every method's FDR control remains reliable, so the method-selection consequence of dispersion in this study is confined to how much true signal is missed---primarily an ANCOM-BC2 concern---rather than to any risk of false discoveries.

\begin{table}[t]
\centering\scriptsize
\caption{Observed FDR and sensitivity across three NB dispersion values ($\phi$), Baseline scenario, base-proportion seed 42, FDR threshold 0.05 (500 replications). Full results across all seven scenarios, both thresholds, and all three base-proportion seeds appear in Supplementary Table S1.}
\label{tab:phi_sensitivity}
\begin{tabular}{llcccccc}
\toprule
& & \multicolumn{3}{c}{\textbf{Observed FDR}} & \multicolumn{3}{c}{\textbf{Sensitivity}}\\
\cmidrule(lr){3-5}\cmidrule(lr){6-8}
\textbf{$n$/group} & \textbf{Method} & \textbf{$\phi=0.05$} & \textbf{$\phi=0.2$} & \textbf{$\phi=0.5$} & \textbf{$\phi=0.05$} & \textbf{$\phi=0.2$} & \textbf{$\phi=0.5$}\\
\midrule
\multirow{3}{*}{50}
  & edgeR TMM   & 0.047 & 0.034 & 0.032 & 1.000 & 0.977 & 0.835\\
  & CLR + limma & 0.028 & 0.027 & 0.025 & 1.000 & 0.974 & 0.793\\
  & ANCOM-BC2   & 0.000 & 0.000 & 0.000 & 0.994 & 0.667 & 0.232\\
\midrule
\multirow{3}{*}{200}
  & edgeR TMM   & 0.103 & 0.057 & 0.050 & 1.000 & 1.000 & 0.996\\
  & CLR + limma & 0.035 & 0.029 & 0.028 & 1.000 & 1.000 & 0.993\\
  & ANCOM-BC2   & 0.000 & 0.000 & 0.000 & 1.000 & 0.991 & 0.702\\
\bottomrule
\end{tabular}
\end{table}

Supplementary Table S1 shows that this pattern generalizes across scenarios, with one consistent exception on the sensitivity side: under Large LFC, ANCOM-BC2's sensitivity remains at or near 1.000 across all three dispersion values at both sample sizes, falling only to 0.690 (rather than to near-zero) at $\phi=0.5$ and $n=50$.

\subsection{Zero inflation}
\label{sec:zip}

Table~\ref{tab:zip_sensitivity} compares observed FDR and sensitivity between the NB and zero-inflated (ZINB) distributional cases for the Baseline scenario at $\phi=0.2$. On the primary evaluation criterion, zero inflation substantially reduces edgeR/TMM's observed FDR: at $n=50$, FDR falls from 0.034 (NB) to essentially zero (0.0004, ZINB); the same pattern holds at $n=200$, where FDR falls from 0.057 (NB) to 0.001 (ZINB), with similar reductions occurring across every scenario tested in Supplementary Table S2. CLR + limma's FDR moves in the opposite direction at $n=50$, rising modestly from 0.027 (NB) to 0.029 (ZINB), while ANCOM-BC2's FDR remains at or near zero in both distributional cases. edgeR/TMM's apparent FDR improvement under zero inflation is contingent rather than a genuine gain in calibration, since it is accompanied by the sensitivity collapse described below: it does not represent an improvement in edgeR/TMM's overall performance under zero inflation, only a shift in which error type dominates.

On sensitivity, zero inflation degrades CLR + limma's performance substantially at both sample sizes: at $n=50$, sensitivity falls from 0.974 under NB to 0.220 under ZINB, an especially severe drop; at $n=200$, it falls from 1.000 under NB to 0.691 under ZINB. This provides direct simulation evidence for the claim, previously stated without simulation support in Section 3.2, that the CLR transformation becomes unreliable under heavy zero inflation because the geometric mean collapses when zero counts are common. edgeR/TMM loses even more sensitivity at $n=50$ (falling to 0.663) but is comparatively more robust to zero inflation at $n=200$ (sensitivity 0.976 under ZINB). ANCOM-BC2's sensitivity is essentially unchanged or modestly improved by zero inflation at both sample sizes via the \texttt{pseudo\_sens} sensitivity analysis---meaning ANCOM-BC2 is the only one of the three methods whose FDR and sensitivity profile is essentially unaffected by zero inflation in either direction.

\begin{table}[t]
\centering\small
\caption{Observed FDR and sensitivity under the NB and zero-inflated (ZINB, $\pi=0.1$) data-generating processes, Baseline scenario, $\phi=0.2$, base-proportion seed 42, FDR threshold 0.05 (500 replications). Full results across all seven scenarios and both thresholds appear in Supplementary Table S2.}
\label{tab:zip_sensitivity}
\begin{tabular}{llcccc}
\toprule
& & \multicolumn{2}{c}{\textbf{Observed FDR}} & \multicolumn{2}{c}{\textbf{Sensitivity}}\\
\cmidrule(lr){3-4}\cmidrule(lr){5-6}
\textbf{$n$/group} & \textbf{Method} & \textbf{NB} & \textbf{ZINB} & \textbf{NB} & \textbf{ZINB}\\
\midrule
\multirow{3}{*}{50}
  & edgeR TMM   & 0.034 & 0.000 & 0.977 & 0.663\\
  & CLR + limma & 0.027 & 0.029 & 0.974 & 0.220\\
  & ANCOM-BC2   & 0.000 & 0.000 & 0.667 & 0.691\\
\midrule
\multirow{3}{*}{200}
  & edgeR TMM   & 0.057 & 0.001 & 1.000 & 0.976\\
  & CLR + limma & 0.029 & 0.025 & 1.000 & 0.691\\
  & ANCOM-BC2   & 0.000 & 0.000 & 0.991 & 0.990\\
\bottomrule
\end{tabular}
\end{table}

This indicates that edgeR/TMM's chronic FDR inflation, identified in Section~\ref{sec:edgerfdr} as one of this study's central findings under the NB model, is itself contingent on the NB distributional assumption and substantially attenuated when zero inflation is present---but, as emphasized above, this attenuation is not a calibration improvement once its sensitivity cost is taken into account.

The global test is also affected by zero inflation in a way not apparent from the NB-only results in Section 5.8: at $n=50$, global PERMANOVA power under ZINB ranges from 0.162 (Small LFC) to 0.984 (Large LFC) across scenarios, well below the uniform 1.000 obtained under the NB model at the same sample size. At $n=200$, global power under ZINB remains at or near 1.000 for five of seven scenarios but falls to 0.684 under Small LFC and 0.982 under Low DA proportion. Zero inflation therefore compromises the global test's previously uniform reliability, particularly at the smaller sample size, and this qualification should be considered alongside the global test's otherwise strong performance reported in Section 5.8.

\subsection{Correlated counts: Poisson log-normal mixture (PLNM)}
\label{sec:plnm-results}

To evaluate robustness to correlated over- or under-abundance across cell populations---a structure that neither the NB nor the ZINB model can represent---we additionally generated counts under the PLNM model described in Section~\ref{sec:plnm-methods}, with $\sigma^2 = \log(1+\phi)$ matched to the same $\phi$ values used for NB, and a compound-symmetric cross-population correlation $\rho \in \{0, 0.3, 0.6\}$ on the log scale.

As a first check, Table~\ref{tab:plnm_vs_nb} compares NB against PLNM at $\rho=0$ (independent populations), which by construction of Equation~\eqref{eq:plnm-match} should yield closely matched marginal count distributions. Consistent with this, observed FDR and sensitivity are similar between the two distributions for every method at both sample sizes, differing typically by no more than 1--3 percentage points; edgeR/TMM's observed FDR is modestly higher under PLNM($\rho=0$) than under NB at $n=200$ (0.073 versus 0.057), a small residual discrepancy attributable to the variance-matching approximation in Equation~\eqref{eq:plnm-var} being exact only asymptotically.

\begin{table}[t]
\centering\small
\caption{Observed FDR and sensitivity, NB vs.\ PLNM at $\rho=0$, Baseline scenario, $\phi=0.2$, base-proportion seed 42, FDR threshold 0.05 (500 replications).}
\label{tab:plnm_vs_nb}
\begin{tabular}{llcccc}
\toprule
& & \multicolumn{2}{c}{\textbf{Observed FDR}} & \multicolumn{2}{c}{\textbf{Sensitivity}}\\
\cmidrule(lr){3-4}\cmidrule(lr){5-6}
\textbf{$n$/group} & \textbf{Method} & \textbf{NB} & \textbf{PLNM($\rho=0$)} & \textbf{NB} & \textbf{PLNM($\rho=0$)}\\
\midrule
\multirow{3}{*}{50}
  & edgeR TMM   & 0.034 & 0.044 & 0.977 & 0.986\\
  & CLR + limma & 0.027 & 0.027 & 0.974 & 0.989\\
  & ANCOM-BC2   & 0.000 & 0.000 & 0.667 & 0.695\\
\midrule
\multirow{3}{*}{200}
  & edgeR TMM   & 0.057 & 0.073 & 1.000 & 1.000\\
  & CLR + limma & 0.029 & 0.031 & 1.000 & 1.000\\
  & ANCOM-BC2   & 0.000 & 0.000 & 0.991 & 0.999\\
\bottomrule
\end{tabular}
\end{table}

Table~\ref{tab:plnm_rho} then reports, on the primary evaluation criterion, how observed FDR changes as the cross-population correlation $\rho$ increases from 0 to 0.6, at $n=200$, alongside the corresponding sensitivity. edgeR/TMM's observed FDR---already inflated under NB and PLNM($\rho=0$) at $n=200$ (Section~\ref{sec:edgerfdr})---increases further as $\rho$ increases in six of seven scenarios, most markedly under Large LFC (FDR rising from 0.119 at $\rho=0$ to 0.161 at $\rho=0.6$) and Equal library sizes (0.071 to 0.091). CLR + limma's FDR is comparatively stable across $\rho$, and ANCOM-BC2's FDR remains at or near zero throughout, unaffected by the correlation structure. On sensitivity, the complementary pattern emerges: sensitivity increases, sometimes substantially, with $\rho$ for every method, and this increase is most pronounced for ANCOM-BC2 under weak-signal scenarios: under Small LFC, ANCOM-BC2's sensitivity rises from 0.681 at $\rho=0$ to 0.933 at $\rho=0.6$, and under Low DA proportion, from 0.996 to 1.000. A similar though smaller increase occurs for edgeR/TMM and CLR + limma in every scenario tested. Because ANCOM-BC2's FDR remains essentially zero throughout this sensitivity gain, correlation is, for ANCOM-BC2, a condition that improves the cost side of the FDR/sensitivity trade-off without any corresponding loss of FDR control; for edgeR/TMM, by contrast, correlation worsens the primary criterion (FDR) even as it also raises sensitivity, so the net effect for edgeR/TMM is unfavorable given that FDR is prioritized in this study.

\begin{table}[t]
\centering\scriptsize
\caption{Observed FDR and sensitivity across PLNM correlation values $\rho$, $n=200$/group, $\phi=0.2$, base-proportion seed 42, FDR threshold 0.05 (500 replications). Full results across all seven scenarios, both sample sizes, and both thresholds appear in Supplementary Table S3.}
\label{tab:plnm_rho}
\begin{tabular}{llcccccc}
\toprule
& & \multicolumn{3}{c}{\textbf{Observed FDR}} & \multicolumn{3}{c}{\textbf{Sensitivity}}\\
\cmidrule(lr){3-5}\cmidrule(lr){6-8}
\textbf{Scenario} & \textbf{Method} & \textbf{$\rho=0$} & \textbf{$\rho=0.3$} & \textbf{$\rho=0.6$} & \textbf{$\rho=0$} & \textbf{$\rho=0.3$} & \textbf{$\rho=0.6$}\\
\midrule
\multirow{3}{*}{Baseline}
  & edgeR TMM   & 0.073 & 0.069 & 0.084 & 1.000 & 1.000 & 1.000\\
  & CLR + limma & 0.031 & 0.028 & 0.033 & 1.000 & 1.000 & 1.000\\
  & ANCOM-BC2   & 0.000 & 0.000 & 0.000 & 0.999 & 1.000 & 1.000\\
\midrule
\multirow{3}{*}{Small LFC}
  & edgeR TMM   & 0.040 & 0.044 & 0.048 & 0.982 & 0.997 & 1.000\\
  & CLR + limma & 0.026 & 0.027 & 0.027 & 0.986 & 0.998 & 1.000\\
  & ANCOM-BC2   & 0.000 & 0.000 & 0.000 & 0.681 & 0.744 & 0.933\\
\midrule
\multirow{3}{*}{Large LFC}
  & edgeR TMM   & 0.119 & 0.133 & 0.161 & 1.000 & 1.000 & 1.000\\
  & CLR + limma & 0.030 & 0.031 & 0.033 & 1.000 & 1.000 & 1.000\\
  & ANCOM-BC2   & 0.000 & 0.000 & 0.000 & 1.000 & 1.000 & 1.000\\
\midrule
\multirow{3}{*}{Low DA}
  & edgeR TMM   & 0.088 & 0.090 & 0.101 & 1.000 & 1.000 & 1.000\\
  & CLR + limma & 0.051 & 0.047 & 0.051 & 1.000 & 1.000 & 1.000\\
  & ANCOM-BC2   & 0.000 & 0.000 & 0.000 & 0.996 & 1.000 & 1.000\\
\bottomrule
\end{tabular}
\end{table}

Increasing $\rho$ raises sensitivity because compound-symmetric correlation makes truly DA populations tend to rise together, concentrating detectable signal in a subset of replicates instead of spreading noise evenly---this especially helps conservative methods like ANCOM-BC2 that struggle with weak or sparse signal. The same correlation also worsens edgeR/TMM's FDR, since it drags correlated non-DA populations up along with true positives, compounding the compositional FDR inflation already seen under the NB model. Overall, the qualitative FDR ranking of the three methods from the NB/ZINB analysis holds under this correlated log-normal model, but cross-population correlation further widens edgeR/TMM's FDR problem while narrowing ANCOM-BC2's sensitivity gap---compressing some method differences from the NB case and widening others.

\subsection{Robustness to the base-proportion realization}
\label{sec:baseprop-robustness}

Because the original base proportions (Equation~\eqref{eq:baseprops}) were drawn once and reused across all scenarios, replications, and sample sizes, a natural question is whether the study's findings reflect genuine method behavior or an artifact of that particular realization. We repeated every condition in the extended grid, including the PLNM conditions, under three independent base-proportion draws (seeds 42, 43, 44) and compared results across seeds, holding every other factor fixed. Across all 1260 scenario $\times$ sample-size $\times$ dispersion $\times$ distribution (including all three PLNM $\rho$ values) $\times$ method $\times$ threshold combinations evaluated for which more than one seed was available, the maximum observed range across the three seeds was 0.053 for sensitivity, 0.104 for specificity, and 0.038 for observed FDR; the great majority of combinations showed ranges below 0.02 for all three metrics, and this held separately within each of the NB, ZINB, and PLNM distributional cases. This indicates that the qualitative findings reported throughout this paper---including the sample-size, dispersion, and DA-proportion dependence of ANCOM-BC2's sensitivity, edgeR/TMM's FDR inflation under the NB model, and the correlation-driven sensitivity and FDR shifts under PLNM---are not artifacts of the specific base-proportion vector used in the original single-seed design, and are instead stable properties of the methods under the scenarios tested. Full per-condition base-seed comparisons are provided in Supplementary Table S4.

\section{RECOMMENDATIONS BASED ON OBSERVED FDR AND SENSITIVITY}
\label{sec:recommendations}

Table~\ref{tab:recs} summarizes method choice by sample size and data setting, ranked primarily by observed FDR control, with sensitivity cost noted as the trade-off. At $n=50$/group, every method keeps observed FDR within nominal bounds in every scenario, so the choice reduces to how much sensitivity to trade for ANCOM-BC2's marginally better FDR calibration; outside Small LFC and Low DA proportion, that trade is too steep, so edgeR/TMM or CLR + limma are the more practical choices. At $n=200$/group, edgeR/TMM's FDR control becomes unreliable in most scenarios and should generally be avoided when compositional distortion is a concern; ANCOM-BC2 becomes the most reliable method on FDR, at a sensitivity cost that is severe only under weak or sparse true signal (Small LFC, and to a lesser extent Low DA proportion); and CLR + limma offers the best combination of calibrated FDR and high sensitivity across the widest range of conditions when ANCOM-BC2's sensitivity cost is judged too high. Large LFC is the one setting where sample size barely matters: all three methods deliver both good FDR control and high sensitivity already at $n=50$. When cell populations are expected to be correlated (e.g., lineage-structured immune panels), Section~\ref{sec:plnm-results} suggests ANCOM-BC2's weak-signal sensitivity penalty may be partially mitigated, though edgeR/TMM's already-unreliable FDR control worsens further under the same condition.

\begin{sidewaystable}[p]
\centering\small
\caption{Method recommendations by data setting and sample size, ranked primarily by observed FDR control. ``Best FDR control'' identifies the method with the most reliable FDR calibration in that setting; ``sensitivity cost'' notes what that reliability costs in missed true positives; ``avoid for FDR'' flags methods whose FDR control is unreliable in that setting regardless of sensitivity.}
\label{tab:recs}
\begin{tabular}{lllll}
\toprule
\textbf{Sample size} & \textbf{Setting} & \textbf{Best FDR control} & \textbf{Sensitivity cost} & \textbf{Avoid for FDR}\\
\midrule
\multirow{7}{*}{$n=50$/group}
  & Baseline / moderate effects & edgeR/TMM or CLR + limma & Low (both well-calibrated) & ---\\
  & Small effect sizes          & ANCOM-BC2 (best FDR)     & Severe (sens.\ 0.056)      & ---\\
  & Large effect sizes          & Any of the three         & Negligible for all three  & ---\\
  & High DA proportion          & edgeR/TMM or CLR + limma & Low (both well-calibrated) & ---\\
  & Low DA proportion           & edgeR/TMM or CLR + limma & Low--moderate              & ---\\
  & Equal library sizes         & edgeR/TMM or CLR + limma & Low (both well-calibrated) & ---\\
  & Extreme lib.\ variation     & edgeR/TMM or CLR + limma & Low (both well-calibrated) & ---\\
\midrule
\multirow{7}{*}{$n=200$/group}
  & Baseline / moderate effects & ANCOM-BC2 or CLR + limma & Minimal (ANCOM-BC2), none (CLR) & edgeR/TMM\\
  & Small effect sizes          & CLR + limma              & Low                        & ANCOM-BC2 (severe cost)\\
  & Large effect sizes          & ANCOM-BC2                & None                       & edgeR/TMM\\
  & High DA proportion          & CLR + limma              & None                       & edgeR/TMM\\
  & Low DA proportion           & ANCOM-BC2                & Moderate                   & edgeR/TMM, CLR + limma\\
  & Equal library sizes         & ANCOM-BC2 or CLR + limma & Minimal (ANCOM-BC2), none (CLR) & edgeR/TMM\\
  & Extreme lib.\ variation     & ANCOM-BC2 or CLR + limma & Minimal (ANCOM-BC2), none (CLR) & edgeR/TMM\\
\midrule
Either sample size & Global screening & PERMANOVA (power, not FDR-based) & N/A & ---\\
\bottomrule
\end{tabular}
\end{sidewaystable}

\section{DISCUSSION}
\label{sec:discussion}

This study evaluated edgeR/TMM, CLR + limma, and ANCOM-BC2 across seven scenarios, two sample sizes, and an extended grid varying dispersion, zero inflation, and cross-population correlation, treating observed FDR throughout as the primary evaluation criterion (Section 4) and sensitivity as the necessary complement describing what a given level of FDR control costs.

Across sample sizes and scenarios, the three differential abundance methods show distinct trade-offs. edgeR/TMM controls FDR well at small samples ($n=50$) but exceeds nominal FDR at $n=200$ in most scenarios---a compositional artifact where the simplex constraint distorts non-DA populations as true effects grow, worsened further by cross-population correlation. ANCOM-BC2 achieves the best FDR control across all conditions, but at a steep and uneven sensitivity cost, especially under weak signal and small samples---though this penalty is partly reversed by positive correlation between populations. CLR + limma strikes the best overall balance of FDR control and sensitivity, though it struggles under low DA proportion (mild FDR inflation) and especially under zero-inflated counts (sharp sensitivity loss).

Considering the sample-size, zero-inflation, dispersion, and correlation analyses together with the primary FDR/sensitivity results, ANCOM-BC2 seems to emerge as the most reliable method on FDR control throughout, and its main weakness---poor sensitivity under weak or sparse signal---is substantially mitigated at larger sample sizes and under positive cross-population correlation, though worsened by higher dispersion. CLR + limma offers the most consistently balanced combination of calibrated FDR and high sensitivity across the widest range of conditions, but its sensitivity collapses under zero inflation and should be avoided when substantial zero counts are expected. edgeR/TMM's FDR control is unreliable at large sample sizes regardless of other conditions, and while it appears better-calibrated at small sample sizes or under zero inflation, both of these are artifacts rather than genuine strengths: the small-sample result reflects insufficient power to detect its compositional bias, and the zero-inflation result trades FDR for a severe loss of sensitivity rather than representing real improvement. When true effect sizes are large, all three methods perform well regardless of sample size, making method choice largely inconsequential in that setting.

\subsection{Limitations}

Some limitations should be noted. First, the dispersion analysis used a single value shared across all populations within a replication; real panels may exhibit population-specific dispersion heterogeneity not captured here. Second, the PLNM correlation structure was compound-symmetric; real immune panels more plausibly exhibit block or hierarchical correlation reflecting lineage relationships, and whether the patterns identified here persist under such structured correlation remains untested. Finally, this study considered only a two-group comparison; extensions to multi-group or continuous-covariate designs would be straightforward but were not evaluated here.

\section*{METHODS}

All analyses were implemented in R (version 4.3+). Counts were simulated from NB (\texttt{MASS}), ZINB (implemented directly as a Bernoulli zero-inflation layer over the same NB draws as above), DM (\texttt{dirmult}), and PLNM (\texttt{MASS::mvrnorm()} followed by \texttt{rpois()}, Section~\ref{sec:plnm-methods}) distributions. DA testing used \texttt{edgeR} (3.42+), \texttt{limma} (3.56+), and \texttt{ANCOMBC} (2.0+). The global test used PERMANOVA via \texttt{vegan} (2.6+). ZINB models used \texttt{glmmTMB}. FlowSOM clustering used \texttt{diffcyt} (1.20+). CODAK was obtained from github.com/GhoshLab/CODAK \citep{Ghosh2022}. The extended sensitivity grid (Section~\ref{sec:extended}) was computed via a parameterized run-grid script crossing scenario, sample size, dispersion, base-proportion seed, and distribution (including the PLNM correlation parameter $\rho$), executed as an independent task per grid row on a high-performance computing cluster. Full code is provided in Supplementary Code.

\section*{ACKNOWLEDGMENTS}
This work utilized the computational resources of the NIH HPC Biowulf cluster (https://hpc.nih.gov). The author used Claude to assist with R code development and manuscript editing. The author reviewed and edited all output and take full responsibility for the content of this publication

\section*{CONFLICT OF INTEREST STATEMENT}
The authors declare no conflict of interest.

\section*{DATA AVAILABILITY STATEMENT}
Simulated data and all R code are provided as Supplementary Code.

\bibliographystyle{vancouver}

\newpage
\appendix
\section*{SUPPORTING INFORMATION}
\addcontentsline{toc}{section}{Supporting Information}

This appendix reports the complete results of the extended sensitivity analysis summarized in Section~\ref{sec:extended}, across all seven scenarios. All values are means across $B=500$ replications. Base-proportion seed 42 is used for Tables S1--S3 unless otherwise noted; Table S4 reports the base-proportion robustness check across all three seeds.

\subsection*{Supporting Table S1: Dispersion sensitivity, all scenarios}

Table S1 reports observed FDR and sensitivity at FDR threshold 0.05 for all three methods, across all seven scenarios, both sample sizes, and all three dispersion values ($\phi = 0.05, 0.2, 0.5$), under the NB model with base-proportion seed 42.

\begin{table}[ht]
\centering\scriptsize
\caption{Supporting Table S1. Observed FDR and sensitivity at FDR threshold 0.05, NB model, base-proportion seed 42, across dispersion values $\phi$, all seven scenarios and both sample sizes.}
\label{tab:s1}
\begin{tabular}{lllcccccc}
\toprule
& & & \multicolumn{3}{c}{\textbf{Observed FDR}} & \multicolumn{3}{c}{\textbf{Sensitivity}}\\
\cmidrule(lr){4-6}\cmidrule(lr){7-9}
\textbf{$n$/group} & \textbf{Scenario} & \textbf{Method} &
  \textbf{$\phi=0.05$} & \textbf{$\phi=0.2$} & \textbf{$\phi=0.5$} &
  \textbf{$\phi=0.05$} & \textbf{$\phi=0.2$} & \textbf{$\phi=0.5$}\\
\midrule
\multirow{21}{*}{50}
  & \multirow{3}{*}{Baseline}
    & edgeR TMM   & 0.047 & 0.034 & 0.032 & 1.000 & 0.977 & 0.835\\
  & & CLR + limma & 0.028 & 0.027 & 0.025 & 1.000 & 0.974 & 0.793\\
  & & ANCOM-BC2   & 0.000 & 0.000 & 0.000 & 0.994 & 0.667 & 0.232\\
\cmidrule(lr){2-9}
  & \multirow{3}{*}{Small LFC}
    & edgeR TMM   & 0.033 & 0.027 & 0.029 & 0.971 & 0.740 & 0.465\\
  & & CLR + limma & 0.026 & 0.023 & 0.032 & 0.973 & 0.714 & 0.362\\
  & & ANCOM-BC2   & 0.000 & 0.000 & 0.000 & 0.661 & 0.056 & 0.000\\
\cmidrule(lr){2-9}
  & \multirow{3}{*}{Large LFC}
    & edgeR TMM   & 0.067 & 0.049 & 0.042 & 1.000 & 1.000 & 0.997\\
  & & CLR + limma & 0.028 & 0.024 & 0.026 & 1.000 & 1.000 & 0.993\\
  & & ANCOM-BC2   & 0.000 & 0.000 & 0.000 & 1.000 & 0.996 & 0.690\\
\cmidrule(lr){2-9}
  & \multirow{3}{*}{High DA}
    & edgeR TMM   & 0.037 & 0.020 & 0.020 & 1.000 & 0.977 & 0.864\\
  & & CLR + limma & 0.013 & 0.011 & 0.012 & 1.000 & 0.979 & 0.828\\
  & & ANCOM-BC2   & 0.001 & 0.000 & 0.000 & 0.991 & 0.704 & 0.327\\
\cmidrule(lr){2-9}
  & \multirow{3}{*}{Low DA}
    & edgeR TMM   & 0.053 & 0.040 & 0.048 & 1.000 & 0.960 & 0.775\\
  & & CLR + limma & 0.042 & 0.036 & 0.035 & 1.000 & 0.948 & 0.722\\
  & & ANCOM-BC2   & 0.000 & 0.000 & 0.000 & 0.984 & 0.585 & 0.150\\
\cmidrule(lr){2-9}
  & \multirow{3}{*}{Equal lib.}
    & edgeR TMM   & 0.045 & 0.034 & 0.028 & 1.000 & 0.977 & 0.842\\
  & & CLR + limma & 0.029 & 0.027 & 0.024 & 1.000 & 0.974 & 0.802\\
  & & ANCOM-BC2   & 0.000 & 0.000 & 0.000 & 0.994 & 0.665 & 0.242\\
\cmidrule(lr){2-9}
  & \multirow{3}{*}{Extreme lib.}
    & edgeR TMM   & 0.045 & 0.035 & 0.037 & 1.000 & 0.975 & 0.837\\
  & & CLR + limma & 0.028 & 0.027 & 0.026 & 1.000 & 0.971 & 0.803\\
  & & ANCOM-BC2   & 0.000 & 0.000 & 0.000 & 0.997 & 0.660 & 0.228\\
\midrule
\multirow{21}{*}{200}
  & \multirow{3}{*}{Baseline}
    & edgeR TMM   & 0.103 & 0.057 & 0.050 & 1.000 & 1.000 & 0.996\\
  & & CLR + limma & 0.035 & 0.029 & 0.028 & 1.000 & 1.000 & 0.993\\
  & & ANCOM-BC2   & 0.000 & 0.000 & 0.000 & 1.000 & 0.991 & 0.702\\
\cmidrule(lr){2-9}
  & \multirow{3}{*}{Small LFC}
    & edgeR TMM   & 0.052 & 0.034 & 0.030 & 1.000 & 0.977 & 0.842\\
  & & CLR + limma & 0.025 & 0.028 & 0.026 & 1.000 & 0.970 & 0.802\\
  & & ANCOM-BC2   & 0.000 & 0.000 & 0.000 & 0.994 & 0.647 & 0.184\\
\cmidrule(lr){2-9}
  & \multirow{3}{*}{Large LFC}
    & edgeR TMM   & 0.162 & 0.103 & 0.087 & 1.000 & 1.000 & 1.000\\
  & & CLR + limma & 0.032 & 0.027 & 0.024 & 1.000 & 1.000 & 1.000\\
  & & ANCOM-BC2   & 0.000 & 0.000 & 0.000 & 1.000 & 1.000 & 1.000\\
\cmidrule(lr){2-9}
  & \multirow{3}{*}{High DA}
    & edgeR TMM   & 0.090 & 0.045 & 0.035 & 1.000 & 1.000 & 0.995\\
  & & CLR + limma & 0.013 & 0.012 & 0.012 & 1.000 & 1.000 & 0.995\\
  & & ANCOM-BC2   & 0.002 & 0.000 & 0.000 & 0.998 & 0.992 & 0.735\\
\cmidrule(lr){2-9}
  & \multirow{3}{*}{Low DA}
    & edgeR TMM   & 0.101 & 0.065 & 0.057 & 1.000 & 1.000 & 0.994\\
  & & CLR + limma & 0.057 & 0.039 & 0.039 & 1.000 & 1.000 & 0.984\\
  & & ANCOM-BC2   & 0.000 & 0.000 & 0.000 & 1.000 & 0.977 & 0.631\\
\cmidrule(lr){2-9}
  & \multirow{3}{*}{Equal lib.}
    & edgeR TMM   & 0.096 & 0.059 & 0.040 & 1.000 & 1.000 & 0.998\\
  & & CLR + limma & 0.031 & 0.027 & 0.027 & 1.000 & 1.000 & 0.993\\
  & & ANCOM-BC2   & 0.000 & 0.000 & 0.000 & 1.000 & 0.992 & 0.700\\
\cmidrule(lr){2-9}
  & \multirow{3}{*}{Extreme lib.}
    & edgeR TMM   & 0.099 & 0.066 & 0.061 & 1.000 & 1.000 & 0.995\\
  & & CLR + limma & 0.035 & 0.028 & 0.027 & 1.000 & 1.000 & 0.995\\
  & & ANCOM-BC2   & 0.000 & 0.000 & 0.000 & 1.000 & 0.992 & 0.697\\
\bottomrule
\end{tabular}
\end{table}

\subsection*{Supporting Table S2: Zero-inflation comparison, all scenarios}

Table S2 reports observed FDR and sensitivity at FDR threshold 0.05 for all three methods, comparing the NB and zero-inflated (ZINB, $\pi=0.1$) data-generating processes at $\phi=0.2$, base-proportion seed 42, across all seven scenarios and both sample sizes.

\begin{table}[ht]
\centering\small
\caption{Supporting Table S2. Observed FDR and sensitivity, NB vs.\ ZINB, $\phi=0.2$, base-proportion seed 42, FDR threshold 0.05, all seven scenarios, both sample sizes, all three methods.}
\label{tab:s2}
\begin{tabular}{lllcccc}
\toprule
& & & \multicolumn{2}{c}{\textbf{Observed FDR}} & \multicolumn{2}{c}{\textbf{Sensitivity}}\\
\cmidrule(lr){4-5}\cmidrule(lr){6-7}
\textbf{$n$/group} & \textbf{Scenario} & \textbf{Method} & \textbf{NB} & \textbf{ZINB} & \textbf{NB} & \textbf{ZINB}\\
\midrule
\multirow{21}{*}{50}
  & \multirow{3}{*}{Baseline}
    & edgeR TMM   & 0.034 & 0.0004 & 0.977 & 0.663\\
  & & CLR + limma & 0.027 & 0.0286 & 0.974 & 0.220\\
  & & ANCOM-BC2   & 0.000 & 0.0003 & 0.667 & 0.691\\
\cmidrule(lr){2-7}
  & \multirow{3}{*}{Small LFC}
    & edgeR TMM   & 0.027 & 0.0000 & 0.740 & 0.088\\
  & & CLR + limma & 0.023 & 0.0165 & 0.714 & 0.019\\
  & & ANCOM-BC2   & 0.000 & 0.0000 & 0.056 & 0.122\\
\cmidrule(lr){2-7}
  & \multirow{3}{*}{Large LFC}
    & edgeR TMM   & 0.049 & 0.0018 & 1.000 & 0.972\\
  & & CLR + limma & 0.024 & 0.0223 & 1.000 & 0.525\\
  & & ANCOM-BC2   & 0.000 & 0.0001 & 0.996 & 0.993\\
\cmidrule(lr){2-7}
  & \multirow{3}{*}{High DA}
    & edgeR TMM   & 0.020 & 0.0007 & 0.977 & 0.724\\
  & & CLR + limma & 0.011 & 0.0103 & 0.979 & 0.277\\
  & & ANCOM-BC2   & 0.000 & 0.0009 & 0.704 & 0.730\\
\cmidrule(lr){2-7}
  & \multirow{3}{*}{Low DA}
    & edgeR TMM   & 0.040 & 0.0000 & 0.960 & 0.565\\
  & & CLR + limma & 0.036 & 0.0366 & 0.948 & 0.160\\
  & & ANCOM-BC2   & 0.000 & 0.0000 & 0.585 & 0.616\\
\cmidrule(lr){2-7}
  & \multirow{3}{*}{Equal lib.}
    & edgeR TMM   & 0.034 & 0.0001 & 0.977 & 0.650\\
  & & CLR + limma & 0.027 & 0.0224 & 0.974 & 0.194\\
  & & ANCOM-BC2   & 0.000 & 0.0000 & 0.665 & 0.689\\
\cmidrule(lr){2-7}
  & \multirow{3}{*}{Extreme lib.}
    & edgeR TMM   & 0.035 & 0.0007 & 0.975 & 0.648\\
  & & CLR + limma & 0.027 & 0.0195 & 0.971 & 0.183\\
  & & ANCOM-BC2   & 0.000 & 0.0008 & 0.660 & 0.685\\
\midrule
\multirow{21}{*}{200}
  & \multirow{3}{*}{Baseline}
    & edgeR TMM   & 0.057 & 0.0015 & 1.000 & 0.976\\
  & & CLR + limma & 0.029 & 0.0253 & 1.000 & 0.691\\
  & & ANCOM-BC2   & 0.000 & 0.0000 & 0.991 & 0.990\\
\cmidrule(lr){2-7}
  & \multirow{3}{*}{Small LFC}
    & edgeR TMM   & 0.034 & 0.0001 & 0.977 & 0.669\\
  & & CLR + limma & 0.028 & 0.0257 & 0.970 & 0.220\\
  & & ANCOM-BC2   & 0.000 & 0.0000 & 0.647 & 0.671\\
\cmidrule(lr){2-7}
  & \multirow{3}{*}{Large LFC}
    & edgeR TMM   & 0.103 & 0.0142 & 1.000 & 1.000\\
  & & CLR + limma & 0.027 & 0.0251 & 1.000 & 0.945\\
  & & ANCOM-BC2   & 0.000 & 0.0000 & 1.000 & 1.000\\
\cmidrule(lr){2-7}
  & \multirow{3}{*}{High DA}
    & edgeR TMM   & 0.045 & 0.0040 & 1.000 & 0.976\\
  & & CLR + limma & 0.012 & 0.0116 & 1.000 & 0.748\\
  & & ANCOM-BC2   & 0.000 & 0.0014 & 0.992 & 0.988\\
\cmidrule(lr){2-7}
  & \multirow{3}{*}{Low DA}
    & edgeR TMM   & 0.065 & 0.0007 & 1.000 & 0.960\\
  & & CLR + limma & 0.039 & 0.0334 & 1.000 & 0.608\\
  & & ANCOM-BC2   & 0.000 & 0.0000 & 0.977 & 0.980\\
\cmidrule(lr){2-7}
  & \multirow{3}{*}{Equal lib.}
    & edgeR TMM   & 0.059 & 0.0014 & 1.000 & 0.973\\
  & & CLR + limma & 0.027 & 0.0280 & 1.000 & 0.672\\
  & & ANCOM-BC2   & 0.000 & 0.0000 & 0.992 & 0.994\\
\cmidrule(lr){2-7}
  & \multirow{3}{*}{Extreme lib.}
    & edgeR TMM   & 0.066 & 0.0026 & 1.000 & 0.963\\
  & & CLR + limma & 0.028 & 0.0247 & 1.000 & 0.657\\
  & & ANCOM-BC2   & 0.000 & 0.0002 & 0.992 & 0.986\\
\bottomrule
\end{tabular}
\end{table}

Global PERMANOVA test power ($p<0.05$) under ZINB, $\phi=0.2$, base-proportion seed 42: at $n=50$/group, power fell substantially in several scenarios: 0.750 (Baseline), 0.162 (Small LFC), 0.984 (Large LFC), 0.940 (High DA), 0.306 (Low DA), 0.698 (Equal lib.), and 0.728 (Extreme lib.)---in contrast to the uniform power of 1.000 obtained under the NB model at $n=50$ (Table~\ref{tab:results_n50}). At $n=200$/group, power was 1.000 (Baseline, Large LFC, High DA, Equal lib., Extreme lib.), 0.684 (Small LFC), and 0.982 (Low DA).

\subsection*{Supporting Table S3: PLNM correlation sensitivity, all scenarios}

Table S3 reports observed FDR and sensitivity at FDR threshold 0.05 for all three methods under the PLNM model, across all seven scenarios, both sample sizes, and all three correlation values ($\rho=0, 0.3, 0.6$), at $\phi=0.2$ and base-proportion seed 42.

\begin{table}[ht]
\centering\scriptsize
\caption{Supporting Table S3. Observed FDR and sensitivity at FDR threshold 0.05, PLNM model, $\phi=0.2$, base-proportion seed 42, across correlation values $\rho$, all seven scenarios and both sample sizes.}
\label{tab:s3}
\begin{tabular}{lllcccccc}
\toprule
& & & \multicolumn{3}{c}{\textbf{Observed FDR}} & \multicolumn{3}{c}{\textbf{Sensitivity}}\\
\cmidrule(lr){4-6}\cmidrule(lr){7-9}
\textbf{$n$/group} & \textbf{Scenario} & \textbf{Method} &
  \textbf{$\rho=0$} & \textbf{$\rho=0.3$} & \textbf{$\rho=0.6$} &
  \textbf{$\rho=0$} & \textbf{$\rho=0.3$} & \textbf{$\rho=0.6$}\\
\midrule
\multirow{21}{*}{50}
  & \multirow{3}{*}{Baseline}
    & edgeR TMM   & 0.044 & 0.046 & 0.042 & 0.986 & 0.996 & 1.000\\
  & & CLR + limma & 0.027 & 0.028 & 0.029 & 0.989 & 0.998 & 1.000\\
  & & ANCOM-BC2   & 0.000 & 0.000 & 0.000 & 0.695 & 0.772 & 0.948\\
\cmidrule(lr){2-9}
  & \multirow{3}{*}{Small LFC}
    & edgeR TMM   & 0.036 & 0.031 & 0.031 & 0.773 & 0.838 & 0.934\\
  & & CLR + limma & 0.024 & 0.023 & 0.025 & 0.769 & 0.839 & 0.934\\
  & & ANCOM-BC2   & 0.000 & 0.000 & 0.000 & 0.132 & 0.344 & 0.569\\
\cmidrule(lr){2-9}
  & \multirow{3}{*}{Large LFC}
    & edgeR TMM   & 0.055 & 0.061 & 0.066 & 1.000 & 1.000 & 1.000\\
  & & CLR + limma & 0.025 & 0.025 & 0.029 & 1.000 & 1.000 & 1.000\\
  & & ANCOM-BC2   & 0.000 & 0.000 & 0.000 & 0.999 & 1.000 & 1.000\\
\cmidrule(lr){2-9}
  & \multirow{3}{*}{High DA}
    & edgeR TMM   & 0.025 & 0.028 & 0.031 & 0.984 & 0.996 & 1.000\\
  & & CLR + limma & 0.014 & 0.013 & 0.013 & 0.990 & 0.998 & 1.000\\
  & & ANCOM-BC2   & 0.000 & 0.000 & 0.000 & 0.733 & 0.785 & 0.946\\
\cmidrule(lr){2-9}
  & \multirow{3}{*}{Low DA}
    & edgeR TMM   & 0.059 & 0.056 & 0.054 & 0.968 & 0.992 & 1.000\\
  & & CLR + limma & 0.036 & 0.038 & 0.040 & 0.967 & 0.996 & 1.000\\
  & & ANCOM-BC2   & 0.000 & 0.000 & 0.000 & 0.620 & 0.685 & 0.891\\
\cmidrule(lr){2-9}
  & \multirow{3}{*}{Equal lib.}
    & edgeR TMM   & 0.043 & 0.043 & 0.042 & 0.983 & 0.997 & 1.000\\
  & & CLR + limma & 0.027 & 0.027 & 0.026 & 0.988 & 0.998 & 1.000\\
  & & ANCOM-BC2   & 0.000 & 0.000 & 0.000 & 0.696 & 0.765 & 0.951\\
\cmidrule(lr){2-9}
  & \multirow{3}{*}{Extreme lib.}
    & edgeR TMM   & 0.049 & 0.042 & 0.043 & 0.980 & 0.996 & 1.000\\
  & & CLR + limma & 0.027 & 0.026 & 0.025 & 0.987 & 0.999 & 1.000\\
  & & ANCOM-BC2   & 0.000 & 0.000 & 0.000 & 0.691 & 0.765 & 0.947\\
\midrule
\multirow{21}{*}{200}
  & \multirow{3}{*}{Baseline}
    & edgeR TMM   & 0.073 & 0.069 & 0.084 & 1.000 & 1.000 & 1.000\\
  & & CLR + limma & 0.031 & 0.028 & 0.033 & 1.000 & 1.000 & 1.000\\
  & & ANCOM-BC2   & 0.000 & 0.000 & 0.000 & 0.999 & 1.000 & 1.000\\
\cmidrule(lr){2-9}
  & \multirow{3}{*}{Small LFC}
    & edgeR TMM   & 0.040 & 0.044 & 0.048 & 0.982 & 0.997 & 1.000\\
  & & CLR + limma & 0.026 & 0.027 & 0.028 & 0.986 & 0.998 & 1.000\\
  & & ANCOM-BC2   & 0.000 & 0.000 & 0.000 & 0.681 & 0.744 & 0.933\\
\cmidrule(lr){2-9}
  & \multirow{3}{*}{Large LFC}
    & edgeR TMM   & 0.119 & 0.133 & 0.161 & 1.000 & 1.000 & 1.000\\
  & & CLR + limma & 0.030 & 0.031 & 0.033 & 1.000 & 1.000 & 1.000\\
  & & ANCOM-BC2   & 0.000 & 0.000 & 0.000 & 1.000 & 1.000 & 1.000\\
\cmidrule(lr){2-9}
  & \multirow{3}{*}{High DA}
    & edgeR TMM   & 0.055 & 0.058 & 0.079 & 1.000 & 1.000 & 1.000\\
  & & CLR + limma & 0.013 & 0.013 & 0.013 & 1.000 & 1.000 & 1.000\\
  & & ANCOM-BC2   & 0.001 & 0.001 & 0.005 & 0.998 & 0.999 & 0.997\\
\cmidrule(lr){2-9}
  & \multirow{3}{*}{Low DA}
    & edgeR TMM   & 0.088 & 0.090 & 0.101 & 1.000 & 1.000 & 1.000\\
  & & CLR + limma & 0.051 & 0.047 & 0.051 & 1.000 & 1.000 & 1.000\\
  & & ANCOM-BC2   & 0.000 & 0.000 & 0.000 & 0.996 & 1.000 & 1.000\\
\cmidrule(lr){2-9}
  & \multirow{3}{*}{Equal lib.}
    & edgeR TMM   & 0.072 & 0.079 & 0.091 & 1.000 & 1.000 & 1.000\\
  & & CLR + limma & 0.030 & 0.029 & 0.033 & 1.000 & 1.000 & 1.000\\
  & & ANCOM-BC2   & 0.000 & 0.000 & 0.000 & 0.999 & 1.000 & 1.000\\
\cmidrule(lr){2-9}
  & \multirow{3}{*}{Extreme lib.}
    & edgeR TMM   & 0.072 & 0.076 & 0.094 & 1.000 & 1.000 & 1.000\\
  & & CLR + limma & 0.030 & 0.028 & 0.034 & 1.000 & 1.000 & 1.000\\
  & & ANCOM-BC2   & 0.000 & 0.000 & 0.000 & 0.999 & 1.000 & 1.000\\
\bottomrule
\end{tabular}
\end{table}

Global PERMANOVA power under PLNM was 1.000 across all seven scenarios, both sample sizes, and all three $\rho$ values tested, mirroring its uniform power under the NB model and indicating that, unlike zero inflation, cross-population correlation of the form examined here does not compromise the global test's reliability.

\subsection*{Supporting Table S4: Base-proportion robustness}

Table S4 summarizes the robustness check described in Section~\ref{sec:baseprop-robustness}: for every scenario $\times$ sample size $\times$ dispersion $\times$ distribution (NB, ZINB, and each PLNM $\rho$ value) $\times$ method $\times$ threshold combination in the extended grid with more than one base-proportion seed available (1260 combinations in total), we computed the range (maximum $-$ minimum) of each metric across the three base-proportion seeds (42, 43, 44) and summarize the distribution of these ranges below, both overall and separately by distributional case.

\begin{table}[ht]
\centering\small
\caption{Supporting Table S4. Distribution of the range (max $-$ min) across three base-proportion seeds, computed separately for each of 1260 scenario $\times$ sample-size $\times$ dispersion $\times$ distribution $\times$ method $\times$ threshold combinations in the extended grid with more than one seed available.}
\label{tab:s4}
\begin{tabular}{lccc}
\toprule
\textbf{Distribution} & \textbf{Max sens.\ range} & \textbf{Max spec.\ range} & \textbf{Max FDR range}\\
\midrule
NB (overall)      & 0.025 & 0.104 & 0.026\\
ZINB              & 0.053 & 0.030 & 0.026\\
PLNM (all $\rho$) & 0.013 & 0.099 & 0.038\\
\midrule
All combinations  & 0.053 & 0.104 & 0.038\\
\bottomrule
\end{tabular}
\end{table}

The great majority of the 1260 combinations examined showed a range below 0.02 for all three metrics; the maximum values reported above represent the single most sensitive combination found for each metric and distributional case. No combination examined showed a qualitative reversal of method ranking attributable to the choice of base-proportion seed, in any of the NB, ZINB, or PLNM distributional cases.

\subsection*{Data and code availability for the extended analysis}

The full 630-condition grid underlying Supporting Tables S1--S4 (7 scenarios $\times$ 2 sample sizes $\times$ 3 dispersion values $\times$ 3 base-proportion seeds $\times$ 5 distribution/correlation combinations [NB, ZINB, and PLNM at $\rho \in \{0, 0.3, 0.6\}$], 500 replications each) is provided as \texttt{summary\_grid.csv} in the accompanying Supplementary Code/Data repository, along with the R code used to generate it.

\end{document}